\documentclass{article}
\usepackage[utf8]{inputenc}

\usepackage{amsmath, amsfonts, amssymb, amsthm}
\usepackage{multirow,multicol,tabularx,booktabs}
\usepackage{graphicx,placeins}
\usepackage[shortlabels]{enumitem}
\usepackage{latexsym}
\usepackage{color}
\usepackage[dvipsnames]{xcolor}
\usepackage[hidelinks]{hyperref}
\usepackage{doi}
\usepackage[ruled]{algorithm2e}

\usepackage{comment} 

\usepackage{bbm, bm, mathtools, verbatim}

\newtheorem{conj}{Conjecture}
\newtheorem{thm}[conj]{\bf Theorem}

\newtheorem{cor}[conj]{\bf Corollary}
\newtheorem{prop}[conj]{\bf Proposition}

\newtheorem{rem}[conj]{\bf Remark}

\usepackage[top=3cm,bottom=3cm,right=3cm,left=2.5cm]{geometry}

\def\bar{\overline}

\def\Ac{{\mathcal A}}

\def\Dc{\mbox{$\mathcal D$}}

\def\Hc{\mbox{$\mathcal H$}}

\def\Nc{\mbox{$\mathcal N$}}

\def\Tc{\mbox{$\mathcal T$}}

\def\Sc{\mbox{$\mathcal S$}}

\def\Rb{{\mathbb R}}

\def\Eb{{\mathbb E}}
\def\Pb{{\mathbb P}}

\def\contiguous{\triangleleft\kern-.20em\triangleright}

\def\1{{\mathbf 1} }

\def\x{ {\bf x}}
\def\a{ {\bf a}}

\def\vv{ {\bf v}}
\def\X{ {\bf X}}

\def\W{ {\bf W}}
\def\V{ {\bf V}}

\def\mds{\medskip}

\newcommand{\var}{{\mathrm Var}}
\newcommand{\Diag}{{\mathrm Diag}}
\newcommand{\pen}{{\mathrm{pen}}}

\newcommand{\btau}{\mbox{\boldmath$\tau$}} 

\begin{document}

\title{Testing for equality between conditional copulas given discretized conditioning events}

\author{Alexis Derumigny\thanks{University of Twente, 5 Drienerlolaan, 7522 NB Enschede, Netherlands. a.f.f.derumigny@utwente.nl},
Jean-David Fermanian\thanks{ENSAE,
5, avenue Henry Le Chatelier, 91764 Palaiseau cedex, France.
jean-david.fermanian@ensae.fr},
Aleksey Min\thanks{Technical University of Munich, Parkring 11, 85748 Garching, Germany.
min@tum.de}}

\date{\today}

\maketitle

\abstract{Several procedures have been recently proposed to test the simplifying assumption for conditional copulas. Instead of considering pointwise
conditioning events, we study the constancy of the conditional dependence structure when some covariates belong to general borelian conditioning subsets. Several test statistics based on the equality of conditional Kendall's tau are introduced, and we derive their asymptotic distributions under the null. When such conditioning events are not fixed ex ante, we propose a data-driven procedure to recursively build such relevant subsets. It is based on decision trees that maximize the differences between the conditional Kendall's taus corresponding to the leaves of the trees. The performances of such tests are illustrated in a simulation experiment. Moreover, a study of the conditional dependence between financial stock returns is managed, given some clustering of their past values. The last application deals with the conditional dependence between coverage amounts in an insurance dataset.
}

\mds


\mds

{\bf MCS 2020:} Primary: 62G10, 62H20; Secondary: 62G20, 62H05.

\mds



\section{Introduction}\label{sect-Introduction}

Copulas express the dependence structures of multivariate random vectors independently of their marginal distributions. Therefore, they often allow insightful two-step model specifications and inference. Copulas have induced many academic papers during the last decades and have become very popular in many applied fields. If several multivariate datasets of the same nature are available, then the natural question that arises is whether their dependence structures coincide. This problem was first tackled in \cite{remillard2009} for the case of two samples. A general testing procedure for several samples has been proposed in \cite{bouzebda2011}.
A similar question arises for conditional dependence structure in multivariate datasets.
The key point is now to state whether some conditional copulas differ or not, given different conditioning events.
This is the purpose of our work. 

\mds

To be specific, assume that we observe $n$ i.i.d. replications $(\X_{i,I}, \X_{i,J}), \, i=1, \dots, n$ of a random vector $\X := (\X_I, \X_J) \in \Rb^d$, where, without loss of generality, the subset of conditioned variables is indexed by $I=\{1, \dots, p\}$ and the subset of conditioning variables by $J = \{p+1, \dots, d\}$, for some integer $p \in \{1,\ldots, d\}$.
For $k \in \{ 1, \dots, p\}$, let $F_{k|J}(\, \cdot \, | \, \X_J=\x_J)$ be the conditional law of $X_k$ given $\X_J=\x_J$, where the conditioning event corresponds to a fixed point $\x_J \in \Rb^{d-p}$.
Following \cite{patton2006a, patton2006b, fermanian2012}, the conditional copula of $\X_I$ given $\X_J = \x_J$, denoted as
$C_{I|J}(\, \cdot \, | \, \X_J=\x_J )$, exists and is defined by an equivalent of Sklar's theorem: for every vector $\x \in \Rb^d$,
\begin{equation*}
    \Pb \big(\X_I \leq \x_I | \X_J =\x_J \big) = C_{I|J}\big( F_{1|J}(x_1 | \X_J=\x_J),\ldots,  F_{p|J}(x_p | \X_J=\x_J) \, \Big| \,  \X_J=\x_J \big),
    \label{cond_cop_def_Patton}
\end{equation*}
with uniqueness of $C_{I|J}(\, \cdot \, | \, \X_J=\x_J )$ when the conditional margins of $\X_I$ given $\X_J = \x_J$ are continuous.
Note that all the maps  $C_{I|J}(\, \cdot \, | \, \X_J = \x_J)$, $\x_J \in \Rb^{d-p}$, are different in general.
Nonetheless, a desirable property would be their constancy w.r.t. the choice of the pointwise conditioning event, in particular for
inference purpose (\cite{hobaek2010}). This is the famous so-called ``simplifying assumption'', that is standard for vine models in particular (\cite{czado2019, kurz}).
It can be written as follow:
\begin{align*}
    \Hc_0 : C_{I|J}(\, \cdot \, | \, \X_J=\x_J) \; \text{does not depend on } \x_J\in \Rb^{d-p}.
\end{align*}
Many tests of this simplifying assumption have appeared in the literature.
A generalized likelihood ratio test was introduced by Acar et al. \cite{acar2013} under a semiparametric framework when the conditional copula belongs to one-dimensional parametric family. Several tests were proposed by Derumigny and Fermanian \cite{DerumignyFermanian2017} in more general nonparametric and semiparametric settings. A recent work by Levi and Craiu \cite{levi2019} consider the problem of establishing the validity of the simplifying assumption in a Bayesian setting. See~\cite{gijbels2015a} too.

\medskip

For $k=1, \dots, p$, let $F_{k|J}(\, \cdot \, | \, \X_J \in A_J)$ be the conditional law of $X_k$ given that $\X_J$ belongs to a borelian subset $A_J$ of $\Rb^{d-p}$ with positive probability.
Similarly to the pointwise case, the conditional copula of $\X_I$ given $\X_J \in A_J$ is defined by
\begin{equation*}
    \Pb \big(\X_I \leq \x_I | \X_J \in A_J \big) = C_{I|J}^{A_J}\big( F_{1|J}(x_1 | \X_J\in A_J),\ldots,  F_{p|J}(x_p | \X_J \in A_J) \, \Big| \,  \X_J \in A_J \big),
    \label{cond_cop_def_0}
\end{equation*}
for every $\x$.
It can be easily checked that the conditional copula $C_{I|J}^{A_J}(\, \cdot \, | \, \X_J \in A_J)$ is the cumulative distribution function of the random vector $\big(F_{1|J}(X_{1}|\X_J\in A_J), \ldots, F_{p|J}(X_{p}|\X_J\in A_J)\big)$ given the event $(\X_J \in A_J)$, when all the latter conditional margins are continuous. This will be assumed in this paper.

\mds

Actually, considering box-type conditioning events provides a natural framework in many situations.
For instance, the analysis of dependencies in some particular area of the
covariate space may be of interest per se.
For instance, bank stress tests will focus on some quantiles of losses, i.e. $A_{k,J}=(\text{Loss} > q_\alpha)$.
When dealing with high level quantiles, it is no longer possible to rely on marginal or joint estimators given pointwise conditioning events (kernel smoothing, e.g.) for semi-parametric models.
Moreover, when the dimension of $\X_J$ is larger than three, discretizing is often the single feasible way of building non- and semiparametric dependence models in practice. Indeed, it is no longer possible to invoke usual nonparametric estimators, due to the curse of dimension.

\mds

We now consider several such sets $A_J$, in order to compare the dependence knowing $(\X_J \in A_J)$, in a similar way as for the usual simplifying assumption.
Therefore, let $\Ac_J := \{A_{1,J},\ldots,A_{m,J}\}$ be a collection of $m$ borelian sets $A_{k,J}$ with probabilities $\Pb (\X_J\in A_{k,J})>0$, $k\in \{1, \ldots, m\}$.
This family $\Ac_J$ may be disjoint or even a partition of $\Rb^{d-p}$, but this is not mandatory in our developments.
As in \cite{DerumignyFermanian2017}, consider the null hypothesis $\bar\Hc_0$ given by
\begin{equation*}
    \bar\Hc_0:
    A_{k,J} \mapsto C_{I|J}^{A_{k,J}} (\, \cdot \, | \X_J \in A_{i,J})
    \, \text{is constant over} \; \Ac_J.
\end{equation*}
Several omnibus tests of $\bar\Hc_0$ have already been proposed in~\cite{DerumignyFermanian2017}. 
All of them are based on empirical counterparts of conditional distributions, with some integration on possibly high-dimensional spaces. This may induce
burdensome numerical problems and slow calculations. To the best of our knowledge, the latter paper is the single one in the literature 
that formally tackles the testing problem of $\bar\Hc_0$.
In this work, we propose a simpler and quicker omnibus test procedure.
It will be related to a less demanding
null hypothesis: the equality of all bivariate conditional Kendall's taus associated to the random vector $\X_I$.

\mds

For the sake of illustration, let us first formulate our null hypothesis when $\X_I$ is a bivariate vector, i.e. $p=2$.
In this case, our null hypothesis $\bar\Hc_0^{\;\tau}$ will be
\begin{equation}\label{Eq-H0-bivariate}
    \bar\Hc_0^{\;\tau}:
    \tau_{1,2|\X_J \in A_{1,J}}
    = \tau_{1,2|\X_J \in A_{2,J}}
    = \ldots = \tau_{1,2|\X_J \in A_{m,J}},
\end{equation}
where $\tau_{1,2|\X_J \in A_{k,J}}$ denotes the Kendall's tau between $X_1$ and $X_2$ given $\X_J \in A_{k,J}$.
In other words, we will test whether the conditional Kendall's taus associated to every subset in $\Ac_J$ are equal.
We recall that
\begin{align*}
    \tau_{1,2|\X_J \in A_{k,J}}
    &:= 4 \int C_{I|J}^{A_{k,J}} (u_1, u_2 | \X_J \in A_{k,J})\, C_{I|J}^{A_{k,J}} (du_1, du_2 | \X_J \in A_{k,J}) -1 \\
    &= \Pb\big( (X_{11}-X_{21})(X_{12}-X_{22}) >0 \, \big| \, \X_{1,J} \in A_{k,J} \, , \, \X_{2,J} \in A_{k,J} \big) \\
    & \quad -
    \Pb\big( (X_{11}-X_{21})(X_{12}-X_{22}) < 0 \, \big| \, \X_{1,J} \in A_{k,J} \, , \, \X_{2,J} \in A_{k,J} \big)\\
    & = 4 p_{1,2|A_{k,J}} - 1, \\
p_{1,2|A_{k,J}} &:=  \Pb\big( X_{11} < X_{21}, X_{12} < X_{22} \, \big| \, \X_{1,J} \in A_{k,J} \, , \, \X_{2,J} \in A_{k,J} \big),
\end{align*}
where $\X_1$ and $\X_2$ are two independent versions of $\X$.
Clearly, $\bar\Hc_0$ implies $\bar\Hc_0^{\;\tau}$ but not the opposite. This reasonable lack of power may be accepted
for the gain in terms of simplicity (implementation, interpretation).

\mds

For the general case $p\geq 2$, let us  consider the Kendall's taus of all possible pairs $(X_i,X_j)$, $(i,j)\in \{1,\ldots,p\}^2$ and $i<j$. For each $A_{k,J} \in \Ac_J$, there are $p(p-1)/2$ conditional Kendall's taus $\tau_{i,j|\X_J \in A_{k,J}}$ and we order them in the  vector $\btau_{I|\X_J \in A_{k,J}}$ given by
\begin{align*}
    \btau_{I|\X_J \in A_{k,J}} :=
    \big(\tau_{1,2|\X_J \in A_{k,J}},
    \tau_{1,3|\X_J \in A_{k,J}}, \ldots,
    \tau_{1,p|\X_J \in A_{k,J}},
    \tau_{2,3|\X_J \in A_{k,J}}, \ldots,
    \tau_{p-1,p|\X_J \in A_{k,J}} \big)'.
\end{align*}
The null hypothesis $\bar \Hc_0^{\;\tau}$ in \eqref{Eq-H0-bivariate} can now be generalized as the new null hypothesis $\bar \Hc_0^{\;\btau}$ given by
\begin{equation*} \label{Eq-H0-multivariate}
    \bar \Hc_0^{\;\btau}:
    \btau_{I|\X_J \in A_{1,J}}
    = \btau_{I|\X_J \in A_{2,J}} = \ldots
    = \btau_{I|\X_J \in A_{m,J}}.
\end{equation*}
In other words, $\bar \Hc_0^{\;\btau} = \cap_{a,b,a\neq b} \bar \Hc_{0}^{\; \tau_{a,b}}$ corresponds to the assumption that the function $A_J \mapsto \btau_{I|\X_J \in A_{J}}$ is constant over $\Ac_J$. Equivalently, this means that for every distinct indices $a, b \in \{1, \dots, p \},$ the function $A_J \mapsto \tau_{a,b|\X_J \in A_{J}}$ is constant over $\Ac_J$.

\mds

Our tests will therefore be based on the equality of each conditional Kendall's taus conditionally to each subset.
The latter quantities are unknown and have to be estimated over the corresponding subsamples, that are of random size.
Inference and hypothesis testing using Kendall's tau is popular in dependence modeling.
This is for example the approach chosen in \cite{Jaser2020} to test symmetry and radial symmetry from two independent samples of equal size. However, note that our framework can deal with more than two subsamples and does not require the equality of their sample sizes.

\begin{rem}
    When the conditioning event is pointwise, a corresponding assumption would be
    \begin{equation*}
        \Hc_0^{\;\btau}:
        \btau_{I|\X_J = \x_J} \; \text{does not depend on } \x_J\in \Rb^{d-p},
    \end{equation*}
    defining $\btau_{I|\X_J = \x_J}$ as the vector stacking all pointwise conditional Kendall's taus for every pair of distinct indices.
    For conditional copulas, it is known that neither of the simplifying assumptions $\Hc_0$ or $\bar \Hc_0$ implies the other one when $\Ac_J$ is finite (see \cite[Section 3.1]{DerumignyFermanian2017}). This means that, unfortunately, there is no relationship between a test of $\Hc_0$ and a test of $\bar\Hc_0$.
    Considering conditional Kendall's tau with pointwise conditioning (corresponding to the null hypothesis $\Hc_0^{\;\btau}$) or subset conditioning (corresponding to the null hypothesis $\bar \Hc_0^{\;\btau}$), a similar phenomenon occurs. In other words, there exists a distribution such that $\bar \Hc_0^{\;\btau}$ is satisfied and $\Hc_0^{\;\btau}$ is not; and there exists another distribution such that $\Hc_0^{\;\btau}$ is satisfied and $\bar \Hc_0^{\;\btau}$ is not.
    For completeness, two such counter-examples are given in Appendix~\ref{sec:no_link_CKT}.
    To fuel intuition, in the case $I=\{1, 2\}$ and denoting by $f_{\X_J}(\cdot)$ the density of $\X_J$ w.r.t. the Lebesgue measure, note that
    \begin{eqnarray*}
    \lefteqn{ p_{1,2 | A}=\Pb\big( X_{11} < X_{21}, X_{12} < X_{22}  | \X_{1,J}\in A, \X_{2,J}\in A   \big)     }\\
    &=& \int_{A\times A} p_{1,2 | \x_J,\x'_J} f_{\X_J}(\x_J) f_{\X_J}(\x'_J)\, d\x_J \, d\x'_J / \Pb(A)^2,
    \end{eqnarray*}
    where, for every $(\x_J,\x'_J)\in \Rb^{d-2}\times  \Rb^{d-2}$,
    $$ p_{1,2 | \x_J,\x'_J}=\Pb\big( X_{11} < X_{21}, X_{12} < X_{22}  | \X_{1,J}=\x_J, \X_{2,J}=\x'_J   \big).$$
    In other words, $p_{1,2 | A}$ is a weighted average of the quantities $p_{1,2 | \x_J,\x'_J}$, when $\x_J$ and $\x'_J$ both describe $A$.
    And this is {\bf not} an average of the pointwise conditional probabilities $p_{1,2 | \x_J}=p_{1,2 | \x_J,\x_J}$ (the latter ones, imposing $\x_J=\x'_J$) that yield $\tau_{1,2|\X_J=\x_J}$.
\end{rem}

\mds

The rest of the manuscript is organized as follows. In Section~\ref{sec:asymptotic-p_2}, we introduce a statistic for testing the hypothesis $\Hc_0^{\;\tau}$ and we state its limit distribution. This construction is generalized for the hypothesis $\Hc_0^{\;\btau}$ in Section~\ref{sec:asymptotic-general_p}.
Section~\ref{bootstrapping} explains how bootstrap schemes can be invoked to calculate p-values.
In Section~\ref{sec:algo_tree}, we propose an algorithm to generate a relevant collection of sets $\Ac_J$. The empirical properties of these methods are discussed in a simulation study (Section~\ref{sec:simulation}). Two applications on real datasets are provided in Section~\ref{sect-applications}: first, the conditional dependence between the S\&P500 and the Eurostoxx is studied in a Copula-GARCH model. Through some properties of the tree of conditioning events that has been built, we highlight some contagion phenomenons across stock indices. The second application focuses on the dependence between different coverages in an insurance dataset.
Proofs have been postponed into the appendix.

\section{Asymptotic test in dimension two}
\label{sec:asymptotic-p_2}

In this section, we study some test statistics for testing the null hypothesis $\bar\Hc_0^{\;\tau}$ and we derive their asymptotic distributions.
A generalization for $p>2$ is based on the same idea and is presented in the next section.

\mds

The hypothesis $\bar\Hc_0^{\;\tau}$ can be rewritten as an $m$-sample problem by defining the subsamples $\Sc_k = \{i=1, \dots, n: \X_{i,J} \in A_{k,J} \}$, for $k \in \{ 1, \dots, m\}$. In this case, we want to test whether the dependence between $X_1$ and $X_2$ is the same across all the samples $\Sc_k$.
Contrary to the usual $m$-sample problem, the sample sizes $N_{k,n}$ are random, as they depend on the unknown law of $\X_J$. In the case of data-driven samples (see Section~\ref{sec:algo_tree}), they even depend on the realizations of $\X$. Moreover, we do not restrict ourselves to disjoint samples. In other words, the conditioning
subsets $A_{k,J}$ may intersect.

\mds

Following~\cite{Derumigny2019kernel} and for any $k\in \{1,\ldots,m\}$, the candidates for being estimators of our conditional Kendall's tau are
\begin{align*}
    \hat \tau_{1,2|\X_J \in A_{k,J} }^{(1)}
    &:= 4 \sum_{i=1}^n \sum_{j=1}^n w^{(k)}_{i,n} w^{(k)}_{j,n}
    \1 \big\{ X_{i,1} < X_{j,1} , X_{i,2} < X_{j,2} \big\} - 1, \displaybreak[0] \\
    \hat \tau_{1,2|\X_J \in A_{k,J}}^{(2)}
    &:= \sum_{i=1}^n \sum_{j=1}^n w^{(k)}_{i,n} w^{(k)}_{j,n}
    \Big( \1 \big\{ (X_{i,1} - X_{j,1}) (X_{i,2} - X_{j,2}) > 0 \big\} \\
    &\hspace{2cm} - \1 \big\{ (X_{i,1} - X_{j,1}) (X_{i,2} - X_{j,2}) < 0 \big\} \Big), \\
    \hat \tau_{1,2|\X_J \in A_{k,J}}^{(3)}
    &:= 1 - 4 \sum_{i=1}^n \sum_{j=1}^n w^{(k)}_{i,n} w^{(k)}_{j,n}
    \1 \big\{ X_{i,1} < X_{j,1} , X_{i,2} > X_{j,2} \big\},
\end{align*}
by setting $w^{(k)}_{i,n}:=\1 \{X_{i,J}\in A_{k,J}\}/ N_{k,n}$,  $i=1,\ldots,n$. In contrast to \cite{Jaser2020}, we do not require equal sample sizes for $k=1, \ldots, m$.

\mds

Setting $s^{(k)}_n:= \sum_{i=1}^n \big(w^{(k)}_{i,n} \big)^2$,
it can be easily proved that
$\hat \tau_{1,2|\X_J \in A_{k,J}}^{(2)}$ belongs to the interval
$[- 1 \, , \, 1 - 2 s^{(k)}_n ]$,
$\hat \tau_{1,2|\X_J \in A_{k,J}}^{(2)}$ in
$[- 1 + s^{(k)}_n \, , \, 1 - s^{(k)}_n ]$,
and $\hat \tau_{1,2|\X_J \in A_{k,J}}^{(3)}$ in
$[- 1 + 2 s^{(k)}_n \, , \, 1 ]$.
Moreover, there exists a direct relationship between these three estimators. Indeed, as noticed in~\cite{Derumigny2019kernel}, check that
\begin{equation*}
    \hat \tau_{1,2|\X_J \in A_{k,J}}^{(1)} + s^{(k)}_n = \hat \tau_{1,2|\X_J \in A_{k,J}}^{(2)} = \hat \tau_{1,2|\X_J \in A_{k,J}}^{(3)} - s^{(k)}_n \; \text{a.s.}
    \label{prop:relationship_hat_tau_i}
\end{equation*}
As a consequence, we can rescale the previous estimators so that the new estimator will has values in the whole interval $[-1, 1]$.
This would yield
\begin{equation}
    \hat\tau_{1,2|\X_J \in A_{k,J}}    := \frac{\hat\tau_{1,2|\X_J \in A_{k,J}}^{(1)}}{1-s_n^{(k)}} + \frac{s_n^{(k)}}{1-s_n^{(k)}}
    = \frac{\hat\tau_{1,2|\X_J \in A_{k,J}}^{(2)}}{1-s_n^{(k)}}
    = \frac{\hat\tau_{1,2|\X_J \in A_{k,J}}^{(3)}}{1-s_n^{(k)}} - \frac{s_n^{(k)}}{1-s_n^{(k)}}\cdot
    \label{rel.between_all_hattau}
\end{equation}
The latter quantity $\hat\tau_{1,2|\X_J \in A_{k,J}} $ will be hereafter our so-called empirical Kendall's tau given $\X_J\in A_{k,J}$ for $k=1, \ldots, m$. Besides, it also coincides with the usual Kendall's tau computed on the subsample $\Sc_k$.

\mds

Under $\bar\Hc_0^{\;\tau}$, all the conditional Kendall's taus $\tau_{1,2|\X_J \in A_{k,J}}$ are the same, and many test statistics could be proposed.
In particular, we would like to build a test of $\bar\Hc_0^{\;\tau}$ based on a random vector whose components are of the type
$$ \Delta_{k,l}:=\sqrt{n} \big( \hat \tau_{1,2|\X_J \in A_{k,J} }  -   \hat \tau_{1,2|\X_J \in A_{l,J} } \big) ,$$
for some $(k,l)$ in $\{1,\ldots,m\}^2$, $k < l$. Since $\Pb(\X_J\in A_{k,J})=\mu_k >0$,  the estimator $\hat \tau_{1,2|\X_J \in A_{k,J} } $ is constructed on a subsample, which is roughly (but not exactly) a fraction $\mu_k$ of the whole sample. Therefore, the random sizes $N_{k,n}$, $k=1, \ldots, m$ will have an influence on the joint limiting law of $\Delta_{k,l}$ for some $(k,l)$ in $\{1,\ldots,m\}^2$, $k < l$.
To this goal, we will directly state the law of the random column vectors
$$
\hat \W_{1,2} := \big( \sqrt{n}( \hat \tau_{1,2|\X_J \in A_{1,J} } - \tau_{1,2|\X_J \in A_{1,J} }) , \ldots, \sqrt{n}( \hat \tau_{1,2|\X_J \in A_{m,J} } - \tau_{1,2|\X_J \in A_{m,J} } ) \big)',\;\text{and}$$
$$
\hat \W_{1,2}^{(j)} := \big( \sqrt{n}( \hat \tau^{(j)}_{1,2|\X_J \in A_{1,J} } - \tau_{1,2|\X_J \in A_{1,J} }) , \ldots, \sqrt{n}( \hat \tau^{(j)}_{1,2|\X_J \in A_{m,J} } - \tau_{1,2|\X_J \in A_{m,J} } ) \big)',
$$
for $j=1,2,3.$ The latter laws will be deduced from the limiting law of
$$ \hat \V := \big( \hat D_1 - D_1,\ldots,\hat D_m - D_m, \hat p_1 - p_1,\ldots,\hat p_m - p_m \big)',$$
where
\begin{eqnarray*}
    \hat D_k &:=& \frac{1}{n(n-1)} \sum_{i=1}^n \sum_{j=1, j\neq i}^n \1\{ X_{i,1}< X_{j,1}, X_{i,2}< X_{j,2},X_{i,J}\in A_{k,J}, X_{j,J}\in A_{k,J} \},\\
    D_k &:=& \Eb[\hat D_k]= \Pb\big( X_{i,1}< X_{j,1}, X_{i,2}< X_{j,2},X_{i,J}\in A_{k,J}, X_{j,J}\in A_{k,J} \big),\\
    \hat p_k&:=&\frac{1}{n}\sum_{i=1}^n \1 \{X_{i,J}\in A_{k,J} \},\;\; p_k=\Pb(X_{i,J}\in A_{k,J}),\; k\in \{1,\ldots,m\}.
\end{eqnarray*}

\mds

Denote by $\Pb_k$ the law of $\X$ given $\X_{J}\in A_{k,J}$, i.e.
$\Pb_k(d\x) = \1\{\x_J\in A_{k,j}\} \, \Pb(d\x)/p_k$. Moreover, set
\begin{eqnarray*}
    \pi(\x_1,\x_2) &:=&\big(\1\{x_{1,1} < x_{2,1},x_{1,2} < x_{2,2}\}+ \1\{x_{2,1} < x_{1,1},x_{2,2} < x_{1,2}\} \big)/2, \\
    I_{k,l} &:=& \int \1\{\x_{3,J}\in A_{k,J}\cap A_{l,J}\} \pi(\x_1,\x_3) \pi(\x_2,\x_3) \,\Pb_k(d\x_1)\,\Pb_l(d\x_2)\,\Pb(d\x_3), \\
    J_{k,l} &:=& \int \1\{\x_{2,J}\in A_{k,J}\cap A_{l,J}\} \pi(\x_1,\x_2) \,\Pb_k(d\x_1)\,\Pb(d\x_2),\; p_{k,l} := \Pb(X_{J}\in A_{k,J}\cap A_{l,J}),
\end{eqnarray*}
for every $k,l\in \{1,\ldots,m\}$. The above notations imply $ D_k= p_k^2 \int \pi(\x_1,\x_2) \, \Pb_k(d\x_1)\, \Pb_k(d\x_2)=p_k J_{k,k}.$
We state  the limiting law of $\hat \V$ in the following theorem, whose proof can be found in Appendix \ref{Proof-Theorem-AN_hatV}.

\mds

\begin{thm}
    \label{AN_hatV}
    When $n$ tends to the infinity, $\sqrt{n}\, \hat \V$ tends in law to a $2m$-dimensional Gaussian random vector $\Nc(0,\Sigma)$, where
    $$  \Sigma := \left[
     \begin{array}{cc}
    \Sigma_{1,1} & \Sigma_{1,2} \\
    \Sigma_{1,2}' & \Sigma_{2,2}
    \end{array} \right],$$
    $$ \Sigma_{1,1}:=[\sigma_{k,l}]_{1\leq k,l \leq m},\;\;\sigma_{k,l}:=4 p_k p_l I_{k,l}  - 4 D_k D_l,$$
    $$\Sigma_{1,2}:=[2 p_k J_{k,l} - 2 D_k p_l]_{1\leq k,l \leq m}, \;\;\Sigma_{2,2}:=[p_{k,l} - p_k p_l]_{1\leq k,l \leq m}.$$
\end{thm}

\mds

Now, let us state the limiting law of $\hat \W_{1,2}^{(j)}$, $j\in \{1,2,3\}$ and $\hat \W_{1,2}$. By virtue of~(\ref{rel.between_all_hattau}) and the relation $s_{n}^{(k)}=1/(n\hat p_k)=O(n^{-1})$,  the four latter statistics have the same limiting law. This asymptotic law is presented in the next proposition, whose proof can be found in Appendix \ref{Proof-Propo-2}.

\medskip

\begin{prop}
    \label{AN_hatW}
    When $n$ tends to the infinity, $\sqrt{n}\,\hat \W_{1,2}$ - or any $\sqrt{n}\hat \W_{1,2}^{(j)}$, $j\in \{1,2,3\}$ - tends in law to a $m$-dimensional Gaussian random vector $\Nc(0,\Delta)$, where $\Delta$ is the matrix
    $$ \Delta := 64\Big[ \frac{ I_{kl}}{p_k p_l} + \frac{ D_k D_l p_{k,l}}{ p_k^3 p_l^3}
    - \frac{ D_l J_{k,l}}{ p_k p_l^3} -  \frac{ D_k J_{l,k}}{ p_l p_k^3} \Big]_{1\leq k,l \leq m}.$$
\end{prop}

\medskip

When the subsets $(A_{k,J})_{k=1,\dots,m}$ are disjoint, we simply get the diagonal matrix
$$\Delta = \Diag(\Delta_k)_{1\leq k \leq m} := 16\,\Diag \Big( 4I_{k,k}/p^2_k - (1 + \tau_{1,2|\X_J \in A_{k,J}})^2/(4p_k) \Big). $$

Note that it is easy to consistently estimate the limiting variance-covariance matrix $\Delta$ by replacing every unknown expectations by their empirical counterparts.
For instance, in the case of disjoint $A_{k,J}$, replace the conditional Kendall's tau by their estimators, $p_k$ by $\hat p_k$ and estimate $I_{k,k}$ by
$$ \hat I_{k,k}:= \frac{1}{n^3 \hat p_k^2} \sum_{i_1,i_2,i_3=1}^n \pi(\X_{i_1},\X_{i_3}) \pi(\X_{i_2},\X_{i_3}) \1\{\X_{i_1,J} \in A_{k,J},\X_{i_2,J} \in A_{k,J},\X_{i_3,J} \in A_{k,J}\}.$$
This yields the estimator
$$\hat\Delta   :=\Diag(\hat\Delta_k) := 16\,\Diag \Big( 4\hat I_{k,k}/\hat p^2_k - (1+ \hat\tau_{1,2|\X_J \in A_{k,J}})^2/(4 \hat p_k) \Big). $$
However, the computational cost of $\hat I_{k,k}$ grows as $O(n^3)$ and can be quite high for large subsample sizes.

\mds

To build a test statistic of $\bar\Hc_0^{\;\tau}$, one can consider a subset $\Sc$ of $q$ couples of indices $(k_i,l_i)$ in $\{1,\ldots,m\}^2$, $k_i\neq l_i$, $i=1,\ldots,q$ and $q<m$.
A $q\times m$ contrast matrix $T$ will describe this set $\Sc$ in the following way: on every line of $T$, say the $i$-th,
all components are zero, except the $k_i$-th and the $l_i$-th ones since $(k_i,l_i)\in \Sc$. And the corresponding elements are $1$ and $-1$.
By construction, we impose that the rank of $T$ is $q$. Then, we can test any set of linear null hypothesis of the form
\begin{align*}
    T  \big(
    \tau_{1,2|X_{J}\in A_{1,J}},
    \tau_{1,2|X_{J}\in A_{2,J}}, \ldots,
    \tau_{1,2|X_{J}\in A_{m,J}} \big)'= {\mathbf 0}_q ,
\end{align*}
where ${\mathbf 0}_q$ is the $q$-dimensional vector consisting of zeros.
For instance, for the null hypothesis $\bar\Hc_0^{\;\tau}$, the contrast  matrix $T$ with rank $m-1$ may be chosen as
\begin{equation} \label{eq:contrast-matrix}
    T := \big[ {\mathbf 1}_{m-1}\ : \  -I_{m-1} \big],
\end{equation}
where ${\mathbf 1}_{m-1}$ is the $(m-1)-$dimensional column vector consisting of ones and $I_m $ is the $(m-1)$ dimensional identity matrix. For $m=4$, this yields
the matrix
\begin{equation*}
    T =
    \begin{bmatrix}
    1 & -1 & 0 & 0\\
    1 & 0 & -1 & 0\\
    1 & 0 & 0 & -1\\
    \end{bmatrix},
    \label{matrix_T}
\end{equation*}
whose rank is equal to $3$. This reduces to testing the zero assumption $\bar\Hc_0^{\;\tau}: \tau_{1,2|X_{J}\in A_{1,J}} = \tau_{1,2|X_{J}\in A_{k,J}}$
for all $k \in \{ 2, \dots, m\}$.
\mds

Due to Proposition~\ref{AN_hatW}, the random vector $\sqrt{n}\,T \hat \W_{1,2}$ asymptotically follows a non-degenerate Gaussian distribution $\Nc(0, T \Delta T')$. Therefore, the following statement holds.

\begin{cor}
    \label{Wald-Stat-dim-2}
    Under the the null hypothesis $\bar\Hc_0^{\;\tau}$,
    $$ \Tc_n:=\hat \W_{1,2}' T'(T \hat\Delta T')^{-1} T \hat \W_{1,2}$$
    tends in law to a chi-squared distributed with $m-1$ degrees of freedom.
\end{cor}

\mds
Note that the columns of the contrast matrix $T$ given in \eqref{eq:contrast-matrix} can be arbitrarily permuted without changing the limiting law of $\Tc_n$ under $\bar\Hc_0^{\;\tau}$.
If the column of ones is the $j-$th column $(j\neq 1)$ of $T$, then the corresponding equivalent formulation of the null assumption is
$$ \bar\Hc_0^{\;\tau}:\,\tau_{1,2|X_{J}\in A_{j,J}} = \tau_{1,2|X_{J}\in A_{k,J}},\; \forall k \in \{ 1, \dots, m \} \backslash \{ j \}.$$
Therefore, without loss of generality, we consider only the contrast matrix \eqref{eq:contrast-matrix} in the sequel.


\section{Asymptotic test for higher dimensions}
\label{sec:asymptotic-general_p}

%

In this section, we are dealing with a $p$-dimensional sub-vector $\X_I$, $p>2$. As before, we would still like to test whether the conditioning subsets $A_{k,J}$, $k\in \{1,\ldots,m\}$ have an influence on the underlying conditional copula of $\X_I$ given $\X_J\in A_{k,J}$ (i.e. Assumption $\bar\Hc_0$).
A natural approach would be to rely on usual bivariate (conditional) Kendall's taus as previously, but by considering all possible pairs $(X_a,X_b)$, $(a,b)\in \{1,\ldots,p\}^2$ and $a<b$.
For the given family $\Ac_J$, it is therefore needed to study the limiting law of the stacked random vectors of interest that are
\begin{eqnarray*}
    \hat \W &:=& \big(\hat \W'_{1,2}, \hat \W'_{1,3},\ldots, \hat \W'_{1,p},\hat \W'_{2,3},\ldots, \hat \W'_{p-1,p} \big)' \; ,
    \text{ or} \\
    \hat \W^{(j)} &:=& \big(\hat \W^{(j)'}_{1,2}, \hat \W^{(j)'}_{1,3},\ldots, \hat \W^{(j)'}_{1,p},\hat \W^{(j)'}_{2,3},\ldots, \hat \W^{(j)'}_{p-1,p} \big)', \; j\in \{1,2,3\},
\end{eqnarray*}
all of them being of size $mp(p-1)/2$. Obviously, we have used the notations
$$
\hat \W_{a,b} := \big( \sqrt{n}( \hat \tau_{a,b|\X_J \in A_{1,J} } - \tau_{a,b|\X_J \in A_{1,J} }) , \ldots, \sqrt{n}( \hat \tau_{a,b|\X_J \in A_{m,J} } - \tau_{a,b|\X_J \in A_{m,J} } ) \big)',\; \text{and}
$$
$$
\hat \W_{a,b}^{(j)} := \big( \sqrt{n}( \hat \tau^{(j)}_{a,b|\X_J \in A_{1,J} } - \tau_{a,b|\X_J \in A_{1,J} }) , \ldots, \sqrt{n}( \hat \tau^{(j)}_{a,b|\X_J \in A_{m,J} } - \tau_{a,b|\X_J \in A_{m,J} } ) \big)',
$$
for every couple of indices $(a,b)$ in $\{1,\ldots,p\}^2$, $a < b$.

\mds

Again, it is necessary to state its limiting law, that will be Gaussian.
And a test of $\cap_{a,b,a\neq b} \Hc_{0}^{\tau_{a,b}} $, and then of $\bar\Hc_0$, will be based on a linear transform of $\tilde\W$.
For example, such a test could be based on the average value of conditional Kendall's tau over all possible couples $(X_a,X_b)$, $a,b=1,\ldots,p$, $a< b$, in the spirit of Kendall and Babington Smith~\cite{Kendall}.

\mds
To derive the asymptotic distribution of $\hat\W$ and $\hat\W^{(j)}$, we will need to generalize Theorem~\ref{AN_hatV}.
Let us first define the following quantities:
\begin{align*}
    \hat \V &:=\big(\hat \V'_{1,2}, \hat \V'_{1,3}, \ldots,
    \hat \V'_{p-1,p}, \hat p_1 - p_1, \ldots, \hat p_m - p_m \big)', \\
    \hat \V_{a,b} &:= \big( \hat D_{a,b,1} - D_{a,b,1},\ldots,\hat D_{a,b,m} - D_{a,b,m} \big)', \\
    \hat D_{a,b,k} &:= \frac{1}{n(n-1)} \sum_{i=1}^n \sum_{j=1, j\neq i}^n \1\{ X_{i,a}< X_{j,a}, X_{i,b}< X_{j,b},X_{i,J}\in A_{k,J}, X_{j,J}\in A_{k,J} \}, \\
    D_{a,b,k} &:= \Eb[\hat D_{a,b,k}]= \Pb\big( X_{i,a} < X_{j,a}, X_{i,b}< X_{j,b},X_{i,J}\in A_{k,J}, X_{j,J}\in A_{k,J} \big),
\end{align*}
for every couple of indices $(a,b)$ in $\{1,\ldots,p\}^2$, $a < b$.
Moreover, set
\begin{align*}
    \pi_{a,b}(\x_1,\x_2) &:=\big(\1\{x_{1,a} < x_{2,a},x_{1,b} < x_{2,b}\}
    + \1\{x_{2,a} < x_{1,a},x_{2,b} < x_{1,b}\} \big)/2, \\
    I_{a,b,a',b',k,l} &:= \int \1\{\x_{3,J}\in A_{k,J}\cap A_{l,J}\}  \pi_{a,b}(\x_1,\x_3) \pi_{a',b'}(\x_2,\x_3) \,\Pb_{k}(d\x_1)\,\Pb_{l}(d\x_2)\,\Pb(d\x_3),\; \text{and}\\
    J_{a,b,k,l} &:= \int \1\{\x_{2,J}\in A_{k,J}\cap A_{l,J}\} \pi_{a,b}(\x_1,\x_2) \,\Pb_k(d\x_1)\,\Pb(d\x_2).
\end{align*}
Note that $ D_{a,b,k,k} = p_k^2 \int \pi_{a,b}(\x_1,\x_2) \, \Pb_{k}(d\x_1)\, \Pb_{k}(d\x_2)=p_k J_{a,b,k,k}.$

\begin{thm}
    \label{AN_hatV_extended}
    When $n$ tends to the infinity, $\sqrt{n}\,\hat \V$ tends in law to a $\big( m p (p-1)/2 + p\big)$-dimensional Gaussian random vector $\Nc(0,\Sigma_e)$. The block matrix $\Sigma_e$ is written as
    $$  \Sigma_e := \left[
    \begin{array}{ccccc}
    \Sigma_{(1,2),(1,2)} & \Sigma_{(1,2),(1,3)} & \ldots & \Sigma_{(1,2),(p-1,p)} & \Sigma_{(1,2),0} \\
    \Sigma_{(1,3),(1,2)} & \Sigma_{(1,3),(1,3)} & \ldots & \Sigma_{(1,3),(p-1,p)} & \Sigma_{(1,3),0} \\
    \vdots & \vdots &\ddots &\vdots &\vdots  \\
    \Sigma_{(p-1,p),(1,2)} & \Sigma_{(p-1,p),(1,3)} & \ldots & \Sigma_{(1,3),(p-1,p)} & \Sigma_{(p-1,p),0} \\
    \Sigma_{0,(1,2)} & \Sigma_{0,(1,3)} & \ldots & \Sigma_{0,(p-1,p)} & \Sigma_{0,0} \\
    \end{array} \right],$$
    where $\Sigma_{(a,b),(a',b')}=\Sigma_{(a',b'),(a,b)}'$ for every couples $(a,b)$ and $(a',b')$.
    Similarly, $\Sigma_{(a,b),0}=\Sigma_{0,(a,b)}'$.
    Moreover, the blockwise components of $\Sigma$ are $m\times m$ matrices given by
    $$ \Sigma_{(a,b),(a',b')}:=[4 p_k p_l I_{a,b,a',b',k,l}  - 4 D_{a,b,k} D_{a',b',l}]_{1 \leq k,l \leq m},$$
    $$\Sigma_{(a,b),0}:=[2 p_k J_{a,b,k,l} - 2 D_{a,b,k} p_l]_{1\leq k,l \leq m},
    \quad \Sigma_{0,0}:=[p_{k,l} - p_k p_l]_{1\leq k,l \leq m}.$$
\end{thm}

\mds

Now, let us state the limiting law of $\hat \W^{(j)}$, $j \in \{1,2,3\}$ and $\hat \W$.
As before, the four latter statistics have the same limiting law.
\begin{prop}
    \label{AN_hatW_extended}
    When $n$ tends to the infinity, $\sqrt{n}\,\hat \W$ (or any $\sqrt{n}\,\hat \W^{(j)}$, $j=1,2,3)$) tends in law to a $mp(p-1)/2$-dimensional Gaussian random vector $\Nc(0,\Delta_e)$, where $\Delta_e$ is the square matrix of size $mp(p-1)/2$ that can be written blockwise as
    $$  \Delta_e := \left[
     \begin{array}{cccc}
    \Delta_{(1,2),(1,2)} & \Delta_{(1,2),(1,3)} & \ldots & \Delta_{(1,2),(p-1,p)}  \\
    \Delta_{(1,3),(1,2)} & \Delta_{(1,3),(1,3)} & \ldots & \Delta_{(1,3),(p-1,p)}  \\
    \vdots & \vdots & \ddots &\vdots   \\
    \Delta_{(p-1,p),(1,2)} & \Delta_{(p-1,p),(1,3)} & \ldots & \Delta_{(p-1,p),(p-1,p)}  \\
    \end{array} \right],$$
    and, for any couples $(a,b)$ and $(a',b')$, $\Delta_{(a,b),(a',b')}$ is the matrix of size $m\times m$ defined by
    $$\Delta_{(a,b),(a',b')} :=
    64\Big[ \frac{ I_{a,b,a',b',k,l}}{p_k p_l} + \frac{ D_{a,b,k} D_{a',b',l} p_{k,l}}{ p_k^3 p_l^3}- 
    \frac{ D_{a',b',l} J_{a,b,k,l}}{ p_k p_l^3} -  \frac{ D_{a,b,k} J_{a',b',l,k}}{ p_l p_k^3} \Big]_{1\leq k,l \leq m}.$$
\end{prop}
In particular, when the subsets $A_{k,J}$ are disjoint, $k\in \{1,\ldots,m\}$, then every submatrix $\Delta_{(a,b),(a',b')}$ is diagonal, and
$$\Delta_{(a,b),(a',b')} :=
    16\,\Diag \Big( 4I_{a,b,a',b',k,k}/p^2_k - (1 + \tau_{a,b|\X_J \in A_{k,J}})(1+ \tau_{a',b'|\X_J \in A_{k,J}})/(4p_k) \Big)_{1\leq k \leq m}, $$
since $4D_{a,b,k}/p_k^2 = 1+ \tau_{a,b|\X_J \in A_{k,J}}$ for every $(a,b)$ and every $k$.

\mds

As above, all the members of $\Delta_e$ can be empirically estimated.
For example, for any $a < b \in \{1, \dots, p \},$ introduce
$$ \hat I_{a,b,a',b',k,k} := \frac{1}{n^3 \hat p_k^2} \sum_{i_1,i_2,i_3} \pi_{a,b}(\X_{i_1},\X_{i_3}) \pi_{a',b'}(\X_{i_2},\X_{i_3}) \1\{\X_{i_1,J} \in A_{k,J},\X_{i_2,J} \in A_{k,J},\X_{i_3,J} \in A_{k,J}\}.$$
Then, in the case of disjoint boxes, we obtain a consistent estimator of $\Delta_{(a,b),(a',b')}$ as
$$\hat\Delta_{(a,b),(a',b')} :=
16\,\Diag \Big( 4 \hat I_{a,b,a',b',k,k}/\hat p^2_k - (1+ \hat\tau_{a,b|\X_J \in A_{k,J}})(1+ \hat\tau_{a',b'|\X_J \in A_{k,J}})/(4 \hat p_k) \Big)_{1\leq k \leq m},$$
providing the estimator $\hat\Delta_e$ of the limiting covariance matrix $\Delta_e$.

\mds

Recalling the matrix $T := \big[ {\mathbf 1}_{m-1}\ : \  -I_{m-1} \big]$, that has rank $(m-1)$.
Let $T_e$ be the $(m-1)p(p-1)/2 \times mp(p-1)/2 $-block-matrix
\begin{equation}
T_e:=I_{p(p-1)/2}\otimes T = \Diag (T,\ldots,T),
\label{del_Te}
\end{equation}
whose rank is $(m-1)p(p-1)/2 $.
Then, $\sqrt{n} T_e \hat \W$ tends in law towards a $\Nc\big(0, T_e \Delta_e T'_e\big)$ under the null hypothesis.
Similarly to Corollary \ref{Wald-Stat-dim-2}, we can build a Wald-type test statistics as follow.
\begin{cor}
    \label{Wald-Stat-gener-dim}
    Under the the null hypothesis $\bar\Hc_{0}^{\;\btau}$, the test statistic
    $\Tc_n^{(e)}:=  n  \hat \W' T_e' \Big( T_e \Delta_e T'_e\Big)^{-1} T_e \hat \W $
    converges in distribution to a chi-squared with $(m-1)p(p-1)/2$ degrees of freedom.
\end{cor}

\section{Bootstrapping and other test statistics}
\label{bootstrapping}
In practice, the covariance matrix $\Delta_e$ can be consistently estimated using U-statistics of order 3. The accuracy of the distributional approximation of the test statistic $\Tc_n^{(e)}$ can suffer from this estimation stage if the sample size $n$ is not large enough. Besides, its computational complexity is $O(n^3)$ for fixed $p$ and $m$.
Therefore, we also consider two different statistics, which do not require the estimation of $\Delta_e$ and whose $p-$values can easily be bootstrapped.

\mds

The first test statistic is based on the maximal absolute deviation between two conditional Kendall's taus over all compared pairs of conditioning subsets $A_{k,J}$ and is defined by
\begin{equation*}
    \Tc_{\infty,n} := | \sqrt{n} T_e \hat \W |_\infty.
\end{equation*}
The second test statistic is defined by
\begin{equation*}
    \Tc_{2,n} :=  n  \hat \W' T_e'  T_e \hat \W,
\end{equation*}
which is the sum of squared differences of two Kendall's taus  over all compared pairs of conditioning subsets $A_{k,J}$.
Under the null hypothesis $\bar \Hc_{0}^{\;\btau}$, the asymptotic distribution of $\Tc_{\infty,n}$ and $\Tc_{2,n}$ are more complex than one of $\Tc_n^{(e)}$ and still depend on the unknown covariance matrix $\Delta_e$.
However, their computation does not require an estimation of the covariance matrix $\Delta_e$ and their asymptotic distribution can quickly be estimated using bootstrap techniques.

\mds

\begin{rem}
    Note that both of these test statistics $\Tc_{\infty,n}$ and $\Tc_{2,n}$ depends on a choice of the contrast matrix~$T_e$. This contrast matrix $T_e$ can be chosen in a random way (such as a random permutation of a given contrast matrix). In this case, random contrast matrices have to be resampled for each bootstrap replication, from the same distribution of contrast matrices. Nevertheless, it would change the limiting law established in Corollaries~\ref{Wald-Stat-dim-2} and~\ref{Wald-Stat-gener-dim}.
    In the following, we prefer to choose a fixed contrast matrix which will be reused for all bootstrapped replications.
\end{rem}

\mds

We will consider two such resampling schemes: Efron's classical bootstrap scheme and the ``conditional boostrap'' scheme proposed by \cite{DerumignyFermanian2017}.
In Efron's classical bootstrap, the bootstrapped sample is obtained by resampling $n$ observations $\X_i^*$ from the initial sample $(\X_1, \dots, \X_n)$ with replacement. The bootstrapped test statistics are respectively defined as
\begin{align*}
    \Tc_{\infty, n}^{**}
    :=|\sqrt{n} T_e \hat \W^* - \sqrt{n} T_e \hat \W |_\infty
    \text{ and }
    \Tc_{2, n}^{**}
    := n  (\hat\W^*-\hat\W)' T_e'  T_e (\hat\W^*-\hat\W),
\end{align*}
where $\hat \W^*$ denotes the bootstrapped statistic $\hat \W$ built on the bootstrapped sample $(\X_1^*,\ldots,\X_n^*)$ (see \cite[p. 378]{VVW}).
The two latter statistics share the same asymptotic distributions as $\Tc_{\infty,n}$ and $\Tc_{2,n}$ respectively under the null hypothesis  $\Hc_{0}^{\; \btau}$, when $n$ tends to the infinity. Their p-values are then computed as the empirical frequency of the events $\{ \Tc_{\infty,n}^{**} > \Tc_{\infty,n}\}$ and $\{ \Tc_{2,n}^{**} > \Tc_{2,n}\}$.

\mds

The conditional bootstrap is defined by the following resampling scheme:
\begin{itemize}
    \item draw $\X_J^*$ among $(\X_{1,J},\ldots,\X_{n,J})$. Denote by $k^*$ the index of the subset $A_{k,J}$ that contains $\X_J^*$. 
    \item draw $\X_I^*$ among the observations that belongs to $A_{k^*,J}$, i.e. along the empirical law of $\X_I$ given $\X_J\in A_{k^*,J}$.
\end{itemize}
We will denote by $\Tc_{\infty, n}^{*}$ and $\Tc_{2, n}^{*}$ the corresponding bootstrapped test statistics.

\section{The building of relevant boxes}
\label{sec:algo_tree}

A practical problem may occur when the dimension $d-p$ of the conditioning random vector $\X_J$ is larger then three or four.
Indeed, except when the boxes are imposed by a particular geometry of the problem or by some specific prior information, it is not obvious to guess what would be the most relevant boxes to be tested. In other words, without knowing whether the dependence between the components of $\X_I$ depend on $\X_J$, it is of interest to build some boxes $A_{1,J},\ldots, A_{m,J}$ so that
the dependence structures of $\X_I$ given $\X_J\in A_{k,J}$ for $k\in \{1,\ldots,m\}$ are the ``most different as possible'' from each other.
This practical problem is particularly relevant in some vine structures for which we would want to weaken the standard simplifying assumption.
In other words, it makes sense to build a realistic vine model for which the dependence copulas of any couple $(X_1,X_2)$ given $\X_J=\x_J$ would not be a constant copula (the usual simplifying assumption) nor a continuous function of $\x_J$ (a difficult task in terms of model specification, in general), but rather an intermediate solution: the latter copulas will be picked up among a finite number of conditional copulas of $(X_1,X_2)$ given $\X_J\in A_{k,J}$, $k\in \{1,\ldots,m\}$.

\mds

To this goal, it will be necessary to build conveniently the latter boxes $A_{k,J}$, $k\in \{1,\ldots,m\}$. Assume $m$ is fixed.
Intuitively, the best sets of boxes will be able to discriminate among the $m$ corresponding conditional copulas in a clear-cut way.
A simple solution is to rely on classification trees.
They allow to build $m$ boxes after successively splitting some components of $\X_J$ into two intervals.
The loss function could be given by a distance $d(\cdot,\cdot)$ between the obtained conditional copulas at every stage.
For instance, set $p=2$ and consider a tree algorithm similar to CART. At the first step, one searches an index $k_1\in \{p+1,\ldots,d\}$ and a threshold $t_1$ so that
$$ (k_1,t_1):=\arg\max_{k,t} d\big( C_{1,2 | X_k \leq t}, C_{1,2 | X_k > t} \big) + \pen(k,t),$$
where the penalty function may be related to the size of the obtained boxes:
for a nonnegative tuning parameter $\alpha$, set $\pen(k,t) = \alpha \min \big( \Pb(X_k >t),\Pb(X_k \leq t) \big)$.
As a variant, we could impose a minimum size $\nu$ for all the boxes by choosing
$$\pen(k,t) = \alpha  \min \big( \Pb(X_k >t),\Pb(X_k \leq t) \big) - M \Big\{  \min \big( \Pb(X_k >t),\Pb(X_k \leq t) \big) < \nu \Big\} ,$$
for some large constant $M>>1$ and a given small $\nu <1$.
In practice, the latter conditional copulas have to be estimated from a $n$-sample of i.i.d. $\X$ realizations. Then, the empirical criterion yields
$$ (k_1,t_1):=\arg\max_{k,t}  d\big( \hat C_{1,2 | X_k \leq t}, \hat C_{1,2 | X_k > t} \big) + \widehat \pen(k,t),$$
and $\widehat \pen (k,t) = \alpha \min \big( \Pb_n(X_k >t),\Pb_n(X_k \leq t) \big)$, with obvious notations and keeping the same name for the solutions. Afterwards, the same procedure is applied on the two subsets of observations, and so on. See Friedman et al. (2001, Section 9) for details.

\mds

The latter procedure may be numerically painful in general, especially due to the inference of the conditional copulas and the calculation of a distance between multivariate cdfs'.
Now, as an alternative, we propose to replace the whole functions $ C_{1,2 | X_k \leq t}$ by some conditional dependence measures, again conditional Kendall's tau.
Indeed, the estimation of a (conditional or not) Kendall's tau is related to a classification task, as noticed in~\cite{Derumigny2019classification}.
The new program would be
$$ (k_1,t_1):=\arg\max_{k,t}  | p_{X_1,X_2 | X_k \leq t} -  p_{X_1,X_2 | X_k > t} |^\gamma + \pen (k,t),$$
$$ p_{X_1,X_2 | X_k \leq t} := \Pb\big( X_{1,1} \leq X_{2,1},X_{1,2} \leq X_{2,2} | X_{1,k}\leq t, X_{2,k}\leq t \big),$$
$$ p_{X_1,X_2 | X_k > t} := \Pb\big( X_{1,1} \leq X_{2,1},X_{1,2} \leq X_{2,2} | X_{1,k}> t, X_{2,k}> t \big),$$
for some $\gamma >0$ and two independent versions $\X_1$ and $\X_2$.
The empirical version of the latter criterion is then
$$ (k_1,t_1):=\arg\max_{k,t}  | \hat p_{X_1,X_2 | X_k \leq t} -  \hat p_{X_1,X_2 | X_k > t} |^\gamma + \widehat\pen (k,t),$$
$$ \hat p_{X_1,X_2 | X_k \leq t} := \frac{1}{n(n-1)} \sum_{i\neq j } \1 \big\{ X_{i,1} \leq X_{j,1},X_{i,2} \leq X_{j,2} | X_{i,k}\leq t, X_{j,k}\leq t \big\},$$
$$ \hat p_{X_1,X_2 | X_k > t} := \frac{1}{n(n-1)} \sum_{i\neq j } \1 \big\{ X_{i,1} \leq X_{j,1},X_{i,2} \leq X_{j,2} | X_{i,k}> t, X_{j,k}> t \big\}.$$
Afterwards, the procedure is going on with the two datasets that are corresponding to the conditioning subsets $(X_{k_1}\leq t_1)$ and $(X_{k_1}> t_1)$ respectively, etc.

\mds

To stop the procedure, several rules can be implemented.
The simplest one it to stop the procedure when the number of obtained categories (boxes) is larger than $m$. When $m$ is even, we can exactly obtain $m$ categories. It is also possible to specify a minimum number of observations for each box, and a minimum difference between the conditional Kendall's taus from two estimated boxes.
A more sophisticated way of working is a ``pruning'' rule, once a large tree has been built,
see \cite[p.270]{friedman2001}.

\mds

When $p > 2$ conditioned variables are available, the first step of our criteria becomes
\begin{align}
    (i_1, j_1, k_1, t_1) := \arg\max_{i,j,k,t}  | p_{X_i,X_j | X_k \leq t} -  p_{X_i,X_j | X_k > t} |^\gamma + \pen (i,j,k,t).
    \label{eq:criteria_split}
\end{align}

The complete algorithm is detailed as Algorithm~\ref{algo:tree_based_CKT}, where we fix $\gamma = 1$ and we use the notation $[a,b]_k := \Rb^{k-p-1} \times [a,b] \times \Rb^{d-k} $ to denote the hyper-rectangle of (conditioning) points $\x_J$ satisfying $x_{k} \in [a,b]$. To sum up, the algorithm consists of one recursive function $\texttt{CutCKT}$ which will choose the best boxes $(-\infty,t]_k$ and $(t,\infty)_k$ that can separate the conditional Kendall's tau between two of the conditioned variables as much as possible. Each of these two sets is then recursively partitioned in the same way until the sample size corresponding to each box is too small, or until the difference in conditional Kendall's taus is too small.

\mds

\begin{algorithm}[tb]
\label{algo:tree_based_CKT}
\SetAlgoLined
\SetAlgoNoEnd
\SetKwFunction{CutCKT}{CutCKT}
\SetKwProg{Fn}{def}{\string:}{}
\linespread{1.35}
\selectfont

\Fn({ }){\CutCKT{a dataset $\Dc \in \Rb^{n \times d}$, a subset $A \overset{default}{=} \Rb^{|J|}$, $\emph{\texttt{minCut}} \geq 0$, $\emph{\texttt{minSize}} \geq 0$}}{
    \For{$i\leftarrow 1$ \KwTo $p-1$}{
        \For{$j\leftarrow i+1$ \KwTo $p$}{
            \For{$k\leftarrow p+1$ \KwTo $d$}{
                \ForEach{$t \in \Rb$}{
                    $\texttt{Diff}[i,j,k,t] \leftarrow
                    \big| \hat \tau_{i,j| \X_J \in A \cap (- \infty, t]_k}
                    - \hat \tau_{i,j| \X_J \in A \cap (t, + \infty)_k} \big|$ \;
                }
            }
        }
    }
    $(i^*, j^*, k^*, t^*) \leftarrow \arg\max_{i,j,k,t} \texttt{Diff}[i,j,k,t]
    $ \;

    $\texttt{Box1} \leftarrow A \cap (- \infty, t^*]_{k^*}$ \;
    $\texttt{Box2} \leftarrow A \cap (t^*, + \infty)_{k^*}$ \;

    \eIf{$\min \big( \Pb_n(X_J \in \emph{\texttt{Box1}})
    \, , \, \Pb_n(X_J \in \emph{\texttt{Box2}}) \big) < \emph{\texttt{minSize}}$ \emph{or}
    $\emph{\texttt{Diff}}[i^*,j^*,k^*,t^*] < \emph{\texttt{minCut}}$}{
    \vspace{0.3em}
        \KwRet{$\big( \hat \btau_{I|\X_J \in A} \big)$.}
    }
    {
        $\texttt{Child}_- \leftarrow$ \CutCKT{$\Dc$,
        $A = \emph{\texttt{Box1}}$, $\emph{\texttt{minCut}}$, $\emph{\texttt{minSize}}$} \;

        $\texttt{Child}_+ \leftarrow$ \CutCKT{$\Dc$,
        $A = \emph{\texttt{Box2}}$, $\emph{\texttt{minCut}}$, $\emph{\texttt{minSize}}$} \;
        \KwRet{$\big( \hat \btau_{I|\X_J \in A}, (i^*, j^*, k^*, t^*),
        \emph{\texttt{Child}}_-, \emph{\texttt{Child}}_+ \big)$}
    }
    \vspace{0.3em}
}
\caption{Recursive algorithm for building a set of relevant boxes for conditional Kendall's taus}
\end{algorithm}

Therefore, the object returned by the function $\texttt{CutCKT}$ is a proper binary tree. Its \texttt{Root} stores the vector of (estimated) unconditional Kendall's tau $\hat \btau_{I}$. If the \texttt{Root} has two children, then it also stores the indexes $(i_1, j_1)$ of the pair of conditioned variables selected as having the maximum difference of conditional Kendall's tau following~(\ref{eq:criteria_split}). In this case, it also stores the index $k_1$ of the conditional variable selected, as well as the threshold $t_1$. Recursively, a \texttt{Child} of any \texttt{Node} of the tree is either a final leaf of the tree, with a conditional Kendall's tau $\hat \btau_{I | \X_J \in A}$ where $A$ is the box corresponding to the conditioning subset passed to the \texttt{Child}. Or it is an internal leaf of the tree, composed of the conditional Kendall's tau $\hat \btau_{I | \X_J \in A}$, the indexes $i^*, j^*, k^*$ and the threshold defining the split, and of two children respectively corresponding to the lower box $A \cap (- \infty, t^*]_{k^*}$ (adding the event $\{ X_{k^*} \leq t^* \}$) and to the upper box $A \cap (t^*, +\infty)_{k^*}$ (adding the event $\{ X_{k^*} > t^* \}$).
The type of such a tree can be recursively defined as
\begin{align*}
    &\texttt{Tree} = \hat \btau_I \, \Big\| \, \Big(\hat \btau_I, i_1, j_1, k_1, t_1, \texttt{Child} \big((- \infty, t_1]_{k_1} \big),
    \texttt{Child} \big((t_1, + \infty)_{k_1} \big) \Big), \\
    &\texttt{Child}(A) = \hat \btau_{I|\X_J \in A} \, \Big\| \, \Big(\hat \btau_{I|\X_J \in A}, i^*,j^*,k^*,t^*,
    \texttt{Child} \big(A \cap (- \infty, t^*]_{k^*} \big),
    \texttt{Child} \big(A \cap (t^*, + \infty)_{k^*} \big) \Big),
\end{align*}
where the symbol $\|$ refers here to the union of types.
Examples of such trees are displayed in Figures~\ref{fig:treeL5} and \ref{fig:treeL5t}.

\section{Simulation studies}
\label{sec:simulation}

Two simulation studies are performed to investigate the empirical level and power of the proposed tests as well as the performance of the procedure for building boxes. The significance level $\alpha$ of the proposed tests is set at $5\%$.

\medskip

In the first simulation study, we will consider two or three dimensional vectors $\X_I$, and a one dimensional conditioning variable $\X_J$ only.
Indeed, if the dimension of $\X_J$ increases then the number of boxes increases exponentially with respect to its dimension and the number of observations for each box decreases at the same rate on average. Nonetheless, we can evaluate the influence of the dimension of $\X_J$ within our simulation study by linearly increasing the number of boxes. Since it is not reasonable to estimate Kendall's tau only with few observations, we aim to have at least 25 observations per box on average in our simulation study. Note that   $\Pb\big( \X_{J} \in A_{k,J} \big)$ has to be positive for any $k=1, \ldots, m$. Moreover, each box must contain at least $2$ observations in order to estimate a corresponding  Kendall's tau. 
This empirical restriction will be satisfied by any  randomly drawn sample as well as by any bootstrap sample.  

\medskip

In the second simulation study, we consider a situation for which the conditioning subsets are unknown. Thus, they will be constructed through the data-driven procedure proposed in Algorithm~\ref{algo:tree_based_CKT}.
Half of the sample is used for building the conditioning subset while the other half is used to compute the test statistic as well as its bootstrapped counterpart. This allows us not to contaminate the computation of the p-value by the information used to construct the tree, ensuring that both parts of the process are independent.
We consider every combination of dimensions where the conditioned vector has a dimension between $2$ and $5$ and the conditioning vector has a dimension between $1$ and~$5$. For each setting, we estimate the level and power of our testing methods.

%

\subsection{Empirical level}
\label{emp_level}
To investigate empirical levels, we consider the four-dimensional random vector $\X$. The conditioning random variable $X_4$ follows a standard normal distribution ($J=\{4\}$).  We employ the theoretical quantiles of $X_4$ to build the subsets $A_{k,J}$, $k\in \{1, \ldots, m\}$. Any partition of the sample space of $X_{4}$ is admissible. For each conditioning set $A_{k,J}$, the conditional distribution of $\X_I$ given $X_4\in A_{k,J}$ will be chosen as a three dimensional Gaussian distribution with different mean vectors but with the same correlation matrix $\Sigma$ for every $k\in \{1,\ldots,m\}$. Without loss of generality, the marginal conditional variances are equal to one.
Therefore, we are under $\bar \Hc_0^{\;\btau}$  because the conditional (Gaussian, here) copula of $\X_I$ given $X_4$ only depends on conditional correlations and not on conditional means and/or variances.

\mds

First, we would like to investigate the influence of  the sample size on the empirical level of our testing procedures based on $\Tc_n^{(e)}$,  $\Tc_{\infty,n}^{**}$,  $\Tc_{2,n}^{**}$, $\Tc_{\infty,n}^*$ and  $\Tc_{2,n}^*$.
To this goal, consider $m=4$ boxes $A_{1,J}:=(-\infty, z_{0.25}]$, $A_{2, J}:=( z_{0.25}, z_{0.50}]$, $A_{3, J}:=(  z_{0.50}, z_{0.75}]$ and $A_{4, J}:=(  z_{0.75}, \infty)$, where $z_\alpha$ is the $\alpha$ quantile of the standard normal distribution. Conditional marginal means are specified in Table \ref{tab-level-cond-means}.

\mds

\begin{table}[thb]
\caption{Conditional marginal means of $\X_I$ for four boxes  for the analysis of empirical levels}\label{tab-level-cond-means}
\begin{center}
\begin{tabular}{|c|c|c|c|c|}\hline
        &   $A_{1,J}$ & $A_{2, J}$  & $A_{3, J}$ & $A_{4, J}$ \\ \hline
Cond. marg. mean of $X_1$  &   0   &   2/3  &   4/3&    2  \\ \hline
Cond. marg. mean of $X_2$  &   0   &  -2/3  &  -4/3&   -2  \\ \hline
Cond. marg. mean of $X_3$  &   1   &  1/3   &  -1/3&    1  \\ \hline
\end{tabular}
\end{center}
\end{table} \FloatBarrier
For all boxes $A_{k, J}$, the conditional correlation matrix $\Sigma$ has equal off-diagonal entries $\rho= 0.7071$ (corresponding to $\tau=0.5$).  Thus,  the conditional correlation matrix of $\X_I$ is  independent of $A_{k, J}$, $k\in \{1,2,3,4\}$. Under $\bar \Hc_0^{\;\btau}$, the asymptotic distribution of $\Tc_n^{(e)}$ is $\chi^2_9$ with 9 degrees of freedom. Based on 1000 randomly drawn samples and using the  contrast matrix $T_e$ defined in~(\ref{del_Te}),  Table \ref{tab-level-4-boxes} presents the empirical level of $\Tc_n^{(e)}$, $\Tc_{\infty,n}^{**}$ (classical bootstrap),  $\Tc_{2,n}^{**}$  (classical bootstrap), $\Tc_{\infty,n}^*$ (conditional bootstrap) and $\Tc_{2,n}^*$ (conditional bootstrap) for different sample sizes. Surprisingly, the empirical level of $\Tc_n^{(e)}$  is quite acceptable  even for the small sample size $n=100$ for which each of four Kendall's taus are estimated using 25 observations on the average. In this simulation study, the four bootstrapped   test statistics   result in slightly more  conservative tests.
\begin{table}[thb]\caption{Empirical levels of   five testing procedures for four boxes and different sample sizes}\label{tab-level-4-boxes}
\begin{center}
\begin{tabular}{|c|c|c|c|c|c|}\hline
$n$   &     $\Tc_n^{(e)}$    & $\Tc_{\infty,n}^{**}$      & $\Tc_{2,n}^{**}$     & $\Tc_{\infty,n}^*$     & $\Tc_{2,n}^*$     \\ \hline
100  &    0.033  & 0.017  & 0.022  & 0.018  & 0.021    \\
250   &   0.035  & 0.031  & 0.033  & 0.029  & 0.028    \\
500   &   0.044  & 0.030  & 0.033  & 0.028  & 0.033    \\
1000  &   0.049  & 0.032  & 0.036  & 0.037  & 0.037    \\ \hline
\end{tabular}
\end{center}
\end{table}

\mds
Next, we investigate the influence of the number of boxes $m$ on the empirical levels. To this end, we fix the sample size $n= 500$. Our testing  procedures are  applied with $m=2, 4, 6, \ldots, 20$ intervals/boxes and the results are presented in Table \ref{tab-level-diff-number-boxes} based on 1000 randomly drawn samples.  For each $m$, the end points of the sub-intervals are computed using the theoretical quantiles of the standard normal distribution at $m+1$ equally spaced probabilities $0, 1/m, \ldots,  (m-1)/m, 1$. The marginal conditional distributions of $\X_I$ are   specified  similarly to the previous simulation study. In particular, conditional mean vectors consist of corresponding $m$ equally spaced points.
Further, we have used the  contrast matrix $T$ from \eqref{eq:contrast-matrix} to construct $T_e$ (c.f.~(\ref{del_Te})), i.e. $T_e=I_{p(p-1)/2}\otimes T$.

\medskip

\begin{table}[thb]\caption{Empirical levels of five testing procedures for different number of boxes and sample size $n= 500$}\label{tab-level-diff-number-boxes}
\begin{center}
\begin{tabular}{|c|c|c|c|c|c|c|c|c|c|c|}\hline
$m$                    & 2    & 4    & 6    &8    &10    &12    &14    &16    &18    &20       \\ \hline
$\Tc_n^{(e)}$          & 0.039    &0.051   & 0.047    &0.060    & 0.098    & 0.112    & 0.119    & 0.146    & 0.210    & 0.253  \\
$\Tc_{\infty,n}^{**}$  & 0.050    &0.042   & 0.032    &0.026    & 0.023    & 0.010    & 0.011    & 0.006    & 0.009    & 0.002  \\
$\Tc_{2,n}^{**}$       & 0.044    &0.055   & 0.027    &0.043    & 0.023    & 0.024    & 0.030    & 0.019    & 0.023    & 0.014  \\
$\Tc_{\infty,n}^*$     & 0.052    &0.043   & 0.033    &0.022    & 0.022    & 0.008    & 0.010    & 0.007    & 0.009    & 0.002  \\
$\Tc_{2,n}^*$          & 0.045    &0.051   & 0.027    &0.043    & 0.023    & 0.020    & 0.028    & 0.018    & 0.024    & 0.014  \\
 \hline
\end{tabular}
\end{center}
\end{table}

As expected and as shown in Table \ref{tab-level-diff-number-boxes}, the testing procedure based on $\Tc_n^{(e)}$ becomes more and more liberal with an increasing number of boxes. Since the number of available bivariate observations for an estimation of each Kendall's tau decreases and the number of Kendall's tau increases with an increasing $m$, our estimators of Kendall's taus as well as their asymptotic covariance matrices become less and less accurate. 
In contrast, the tests based on the bootstrap procedures are conservative  for large number of boxes. A possible explanation for this phenomenon could be that an inaccurate estimation of Kendall's tau also takes place for bootstrap samples while the Wald type statistic is compared to a theoretical quantile of a limiting distribution.

\mds

\subsection{Empirical power}

First, we consider a four-dimensional random vector $\X$ and recall the simulation setup of Section~\ref{emp_level} with four boxes. 
Now, assume that the conditional correlation matrix of $\X_I$ consists of equal off-diagonal entries $\rho_k$ that
depend on the box $A_{k,J}$, $k\in \{1,2,3,4\}$. The values of these conditional correlations are deduced from equally spaced Kendall's tau in $[0,0.5]$ and  given in Table \ref{tab-power-cond-corr}. Since these values are different, $\bar \Hc_0^{\;\btau}$ is not satisfied here.
\begin{table}[!t]\caption{Conditional correlations of $\X_I$ for four boxes for the power analysis}\label{tab-power-cond-corr}
\begin{center}
\begin{tabular}{|c|c|c|c|c|}\hline
        &   $A_{1, J}$ & $A_{2, J}$  & $A_{3, J}$ & $A_{4, J}$ \\ \hline
Conditional correlation $\rho$  &   0   &   0.259  &   0.500&    0.707  \\ \hline
Conditional Kendall's tau $\tau$  &   0   &   1/6  &   2/6&    3/6 \\ \hline
\end{tabular}
\end{center}
\end{table}
The empirical powers for different sample sizes are given in Table \ref{tab-power-four-boxes} based on 1000 drawn samples and using the  contrast matrix $T_e$. For a sample size $n=100$, the empirical power of the proposed test procedures is at least  $0.676$, which indicates a good performance of the proposed test procedures. Further, the tests based on $\Tc_{2,n}^{**}$ and $\Tc_{2,n}^{*}$ are slightly more powerful then the ones based  on  $\Tc_{\infty,n}^{**}$  and $\Tc_{\infty,n}^{*}$. Thus, our statistical tests are powerful even for small sample sizes. 
From $n=250$, all the proposed tests have an empirical power virtually close to one.
%
\begin{table}[!t]\caption{Empirical powers of four testing procedures for four boxes and different sample sizes}\label{tab-power-four-boxes}
\begin{center}
\begin{tabular}{|c|c|c|c|c|c|}\hline
$n$   &     $\Tc_n^{(e)}$    & $\Tc_{\infty,n}^{**}$      & $\Tc_{2,n}^{**}$     & $\Tc_{\infty,n}^*$     & $\Tc_{2,n}^*$       \\ \hline
 100  &    0.753  & 0.676  & 0.809  & 0.679  & 0.807    \\
250   &    0.999  & 0.999  & 1.000  & 0.999  & 1.000    \\
500   &    1.000  & 1.000  & 1.000  & 1.000  & 1.000    \\
1000  &    1.000  & 1.000  & 1.000  & 1.000  & 1.000    \\ \hline
\end{tabular}
\end{center}
\end{table}

\mds

Similarly to the analysis for levels, we investigate the influence of the number of boxes $m$ on the empirical power. For this purpose, fix the sample size $n=500$ and $m$ conditional correlations are chosen as before. Our testing  procedures are  applied with $m \in\{ 2, 4, 6, \ldots, 20\}$ intervals/boxes and the results are presented in Table \ref{tab-power-diff-number-boxes} based on 1000 drawn samples and using the  contrast matrix $T_e$. All our considered test statistics perfectly detect the differences in terms of Kendall's taus for  a small number of boxes due to   a sufficient sample size for their reliable estimation.   Note that $\Tc_n^{(e)}$ is  sufficiently powerful even for 20 boxes since it does not keep its significance level of $5\%$ at all. The empirical power of the bootstrapped test statistics decreases starting  from 8 boxes. Due to their conservativity, they are essentially less powerful for $m=14, 16, \ldots, 20$ than the Wald-type statistics. 
Similarly to Table~\ref{tab-power-four-boxes}, the tests based on $\Tc_{2,n}^{**}$ and $\Tc_{2,n}^{*}$ are  more powerful then ones based  on  $\Tc_{\infty,n}^{**}$  and $\Tc_{\infty,n}^{*}$. This is due to the contrast matrix $T_e$, which captures the maximal distinction of Kendall's tau differences.  

\mds

\begin{table}[!h]\caption{Empirical power  of four testing procedures for different number of boxes and sample size $n= 500$}\label{tab-power-diff-number-boxes}
\begin{center}
\begin{tabular}{|c|c|c|c|c|c|c|c|c|c|c|}\hline
$m$                    & 2    & 4    & 6    &8    &10    &12    &14    &16    &18    &20       \\ \hline
$\Tc_n^{(e)}$          & 1.000    & 1.000  & 1.000    &1.000     &0.999     &1.000     & 0.998    & 0.998    & 0.997    &0.999   \\
$\Tc_{\infty,n}^{**}$  & 1.000    & 1.000  & 1.000    &0.999     &0.979     &0.914     & 0.800    & 0.625    & 0.513    &0.343   \\
$\Tc_{2,n}^{**}$       & 1.000    & 1.000  & 1.000    &1.000     &0.995     &0.977     & 0.934    & 0.878    & 0.808    &0.715   \\
$\Tc_{\infty,n}^*$     & 1.000    & 1.000  & 1.000    &1.000     &0.978     &0.913     & 0.787    & 0.628    & 0.500    &0.333   \\
$\Tc_{2,n}^*$          & 1.000    & 1.000  & 1.000    &1.000     &0.995     &0.972     & 0.931    & 0.878    & 0.809    &0.713   \\
 \hline
\end{tabular}
\end{center}
\end{table}

\mds

Now, consider a three-dimensional random vector $\X=(X_1, X_2, X_3)$ with uniformly distributed margins on $[0,1]$.  Further, the distribution of $X_1$ and $X_2$ is a Clayton copula whose parameter $\theta$ depends on the value $x_3$  of the random variable $X_3$. For some fixed parameter $\lambda>0$, this parameter $\theta$ is given by
$$
\theta = \theta(x_3) =  \left\{
\begin{array}{ll}
1 & \,\text{when } x_3 \leq  \lambda  \ ,  \\
5 & \,\text{when } x_3 >  \lambda \ . \\
\end{array}
\right.
$$
Thus, there is a structural break on the parameter $\theta$ and we would like to identify this phenomenon. 
To do this, we fix the sample size $n= 500$ and consider $m$ boxes that are determined by the quantiles of $X_3$, corresponding to equally spaced probabilities $0, 1/m, 2/m, \ldots, 1$. In applications, the point of the structural break is not known. To fix the ideas (and the contrast matrix $T_{e}$), we use the first box as the reference category and compare its Kendall' tau to the remaining ones. Thus, the matrix $T_e$ from \eqref{del_Te} is considered.

\mds

First, we choose   $\lambda \in \{0.01, 0.05, 0.10, 1/6, 0.20, 0.25,  0.50\}$ and draw 1000  three-dimensional copula samples, where the third margin determines the copula parameter of the first two margins. Then we apply our testing procedures with $m$ boxes for $m\in \{2, 4,  \ldots, 20\}$. For each $m$, all sub-intervals/boxes contain  the same amount of observations on average.  Table \ref{tab-power-lambda-m} presents the empirical powers of the tests based on  the statistics $\Tc_n^{(e)}$, $\Tc_{\infty,n}^{**}$, $\Tc_{2,n}^{**}$, $\Tc_{\infty,n}^*$ and $\Tc_{2,n}^*$ to identify the structural break. If $\lambda$ is very small  then only few observations are affected by the structural break. Therefore, it is almost impossible to detect the underlying change in the  parameter of the bivariate Clayton copula.   In Table  \ref{tab-power-lambda-m}, this occurs when $\lambda=0.01$. If more data are influenced by the structural break ($\lambda=0.05$ or $\lambda=0.10$), then 
the identification of this structural change  becomes easier. For  $\lambda>0.10$, the structural break in the parameter is obviously identified by our testing procedures for some values of $m$.

A large number of the boxes makes the length of the intervals $A_{k, J}$ smaller, which becomes comparable with the length of the interval $[0, \lambda]$. Therefore, the best power is expected around $m=1/\lambda$. For all considered test statistics, this can explicitly be observed for $\lambda=1/6$. 
When $\lambda<0.50$, the tests based on $\Tc_{2,n}^{**}$ and $\Tc_{2,n}^*$ are usually much more  powerful than the ones based on $\Tc_{\infty,n}^{**}$ and   $\Tc_{\infty,n}^*$. This can be  explained by our choice of the contrast matrix $T_e$, which compares the most different Kendall's tau of the first box to the remaining  $m-1$ ones. If a reference box is differently chosen,  e.g. in the middle of the interval $[0,1]$,  and corresponds to an ``average behavior'' for data coming from a Clayton copula with parameter $\theta=5$, then only a few large differences between Kendall's taus would be be observed. These numerous small differences between our Kendall's taus would largely contribute to the final value of $\Tc_{2,n}^{**}$ and $\Tc_{2,n}^*$, softening their power. This explains why we have observed that the tests based on  $\Tc_{\infty,n}^{**}$ and   $\Tc_{\infty,n}^*$ are sometimes more powerful than the latter ones. Further, the empirical powers of the test statistics $\Tc_{2,n}^{**}$ and $\Tc_{2,n}^*$ for $\lambda=0.5$ in Table  \ref{tab-power-lambda-m} are not considerably better than the ones of $\Tc_{\infty,n}^{**}$ and $\Tc_{\infty,n}^*$, because of a maximum level of heterogeneity between the dependencies across the chosen boxes.


\begin{table}[!h]\caption{Empirical powers for the bivariate Clayton copula with a structural break in its parameter for different $\lambda$ and $m$ based on 1000 drawn samples.  The sample size $n$ is equal to 500.}\label{tab-power-lambda-m}
\begin{center}
{\small
\begin{tabular}{|c|c|c|c|c|c|c|c|c|c|c|c|}\hline
$\lambda=0.01$   &$m$                     & 2        & 4      &  6       & 8        & 10       &  12        &  14        & 16         &  18        &  20       \\ \cline{2-12}
                 & $\Tc_n^{(e)}$          & 0.019    & 0.019  &  0.019   & 0.012    & 0.017    &  0.024     &  0.014     & 0.010      &  0.011     &  0.016   \\
                 & $\Tc_{\infty,n}^{**}$  & 0.033    & 0.033  &  0.027   & 0.022    & 0.019    &  0.018     &  0.010     & 0.011      &  0.009     &  0.004   \\
                 & $\Tc_{2,n}^{**}$       & 0.033    & 0.039  &  0.036   & 0.040    & 0.049    &  0.045     &  0.045     & 0.046      &  0.045     &  0.041   \\
                 & $\Tc_{\infty,n}^*$     & 0.033    & 0.034  &  0.027   & 0.023    & 0.017    &  0.015     &  0.012     & 0.010      &  0.010     &  0.004   \\
                 & $\Tc_{2,n}^*$          & 0.033    & 0.041  &  0.038   & 0.043    & 0.043    &  0.044     &  0.045     & 0.046      &  0.047     &  0.040   \\      \hline
$\lambda=0.05$   &$m$                     & 2        & 4      &  6       & 8        & 10       &  12        &  14        & 16         &  18        &  20       \\ \cline{2-12}
                 & $\Tc_n^{(e)}$          & 0.165    & 0.227  &  0.227   & 0.210    & 0.222    &  0.213     &  0.229     & 0.214      &  0.205     &  0.207   \\
                 & $\Tc_{\infty,n}^{**}$  & 0.233    & 0.353  &  0.416   & 0.466    & 0.488    &  0.486     &  0.509     & 0.480      &  0.456     &  0.391   \\
                 & $\Tc_{2,n}^{**}$       & 0.233    & 0.408  &  0.517   & 0.598    & 0.663    &  0.689     &  0.718     & 0.728      &  0.747     &  0.723   \\
                 & $\Tc_{\infty,n}^*$     & 0.219    & 0.352  &  0.413   & 0.467    & 0.488    &  0.487     &  0.497     & 0.480      &  0.449     &  0.398   \\
                 & $\Tc_{2,n}^*$          & 0.219    & 0.417  &  0.520   & 0.597    & 0.652    &  0.686     &  0.719     & 0.728      &  0.748     &  0.715   \\      \hline
$\lambda=0.10$   &$m$                     & 2        & 4      &  6       & 8        & 10       &  12        &  14        & 16         &  18        &  20       \\ \cline{2-12}
                 & $\Tc_n^{(e)}$          & 0.549    & 0.730  &  0.806   & 0.827    & 0.778    &  0.706     &  0.640     & 0.610      &  0.559     &  0.534   \\
                 & $\Tc_{\infty,n}^{**}$  & 0.606    & 0.833  &  0.939   & 0.956    & 0.965    &  0.911     &  0.823     & 0.705      &  0.569     &  0.448   \\
                 & $\Tc_{2,n}^{**}$       & 0.606    & 0.885  &  0.966   & 0.990    & 0.989    &  0.977     &  0.936     & 0.895      &  0.819     &  0.749   \\
                 & $\Tc_{\infty,n}^*$     & 0.613    & 0.830  &  0.939   & 0.960    & 0.967    &  0.912     &  0.816     & 0.702      &  0.578     &  0.440   \\
                 & $\Tc_{2,n}^*$          & 0.613    & 0.884  &  0.964   & 0.990    & 0.990    &  0.975     &  0.930     & 0.896      &  0.823     &  0.751   \\      \hline
$\lambda=1/6 $   &$m$                     & 2        & 4      &  6       & 8        & 10       &  12        &  14        & 16         &  18        &  20       \\ \cline{2-12}
                 & $\Tc_n^{(e)}$          & 0.911    & 0.990  &  0.998   & 0.982    & 0.969    &  0.959     &  0.936     & 0.910      &  0.880     &  0.857   \\
                 & $\Tc_{\infty,n}^{**}$  & 0.933    & 0.998  &  1.000   & 0.994    & 0.962    &  0.884     &  0.758     & 0.629      &  0.479     &  0.354   \\
                 & $\Tc_{2,n}^{**}$       & 0.933    & 1.000  &  1.000   & 0.999    & 0.995    &  0.979     &  0.940     & 0.892      &  0.802     &  0.721   \\
                 & $\Tc_{\infty,n}^*$     & 0.930    & 0.998  &  1.000   & 0.996    & 0.963    &  0.889     &  0.763     & 0.644      &  0.483     &  0.353   \\
                 & $\Tc_{2,n}^*$          & 0.930    & 1.000  &  1.000   & 1.000    & 0.997    &  0.977     &  0.944     & 0.882      &  0.794     &  0.704   \\      \hline
$\lambda=0.20$   &$m$                     & 2        & 4      &  6       & 8        & 10       &  12        &  14        & 16         &  18        &  20       \\ \cline{2-12}
                 & $\Tc_n^{(e)}$          & 0.977    & 0.997  &  0.998   & 0.997    & 0.991    &  0.988     &  0.983     & 0.973      &  0.958     &  0.949   \\
                 & $\Tc_{\infty,n}^{**}$  & 0.980    & 1.000  &  1.000   & 0.994    & 0.970    &  0.882     &  0.770     & 0.628      &  0.447     &  0.330   \\
                 & $\Tc_{2,n}^{**}$       & 0.980    & 1.000  &  1.000   & 0.999    & 0.995    &  0.973     &  0.935     & 0.879      &  0.784     &  0.698   \\
                 & $\Tc_{\infty,n}^*$     & 0.985    & 0.999  &  1.000   & 0.993    & 0.971    &  0.889     &  0.765     & 0.633      &  0.450     &  0.332   \\
                 & $\Tc_{2,n}^*$          & 0.985    & 1.000  &  1.000   & 1.000    & 0.996    &  0.976     &  0.938     & 0.875      &  0.784     &  0.700   \\      \hline
$\lambda=0.25$   &$m$                     & 2        & 4      &  6       & 8        & 10       &  12        &  14        & 16         &  18        &  20       \\ \cline{2-12}
                 & $\Tc_n^{(e)}$          & 0.993    & 1.000  &  0.998   & 1.000    & 0.998    &  0.998     &  0.997     & 0.992      &  0.990     &  0.978   \\
                 & $\Tc_{\infty,n}^{**}$  & 0.995    & 1.000  &  1.000   & 0.995    & 0.949    &  0.845     &  0.719     & 0.564      &  0.456     &  0.317   \\
                 & $\Tc_{2,n}^{**}$       & 0.995    & 1.000  &  1.000   & 0.999    & 0.994    &  0.977     &  0.923     & 0.852      &  0.758     &  0.663   \\
                 & $\Tc_{\infty,n}^*$     & 0.994    & 1.000  &  0.999   & 0.994    & 0.949    &  0.846     &  0.725     & 0.570      &  0.449     &  0.316   \\
                 & $\Tc_{2,n}^*$          & 0.994    & 1.000  &  1.000   & 0.999    & 0.994    &  0.975     &  0.927     & 0.858      &  0.765     &  0.664   \\      \hline
$\lambda=0.50$   &$m$                     & 2        & 4      &  6       & 8        & 10       &  12        &  14        & 16         &  18        &  20       \\ \cline{2-12}
                 & $\Tc_n^{(e)}$          & 1.000    & 1.000  &  1.000   & 1.000    & 1.000    &  1.000     &  1.000     & 1.000      &  1.000     &  1.000   \\
                 & $\Tc_{\infty,n}^{**}$  & 1.000    & 1.000  &  0.997   & 0.971    & 0.892    &  0.739     &  0.588     & 0.451      &  0.321     &  0.222   \\
                 & $\Tc_{2,n}^{**}$       & 1.000    & 1.000  &  1.000   & 0.993    & 0.968    &  0.900     &  0.773     & 0.633      &  0.496     &  0.370   \\
                 & $\Tc_{\infty,n}^*$     & 1.000    & 1.000  &  0.996   & 0.971    & 0.895    &  0.739     &  0.581     & 0.441      &  0.322     &  0.230   \\
                 & $\Tc_{2,n}^*$          & 1.000    & 1.000  &  1.000   & 0.993    & 0.967    &  0.901     &  0.778     & 0.633      &  0.482     &  0.372   \\      \hline
\end{tabular}
}
\end{center}
\end{table}

\subsection{Simulations with data-driven boxes}

\medskip

In this section, the ``boxes'' $A_{k,J}$ are not fixed ex ante. 
We will use a sample size $n = 1000$ of observations from $\X$.
We choose the conditioning vector $\X_J$ to be always a $d-p$-dimensional standard Gaussian random vector.
For the level analysis, the conditioning vector $\X_I$ is simulated from a Gaussian copula with all Kendall's taus constant equal to $0$, independently of $\X_J$.

\medskip

In the power analysis, we consider the following conditional distribution for the conditioned vector $\X_I$ given $\X_J$. The conditional copula of $\X_I$ given $\X_J$ is represented by a $d-p$ dimensional D-vine that is composed of bivariate conditional Gaussian copulas. Such copulas are parameterized by their conditional Kendall's tau given $X_{p+1}$
and they satisfy the simplifying assumption, for every fixed value of $X_{p+1}$.
Therefore, we choose $\tau_{1,2|\X_J} = 0.7 - 0.6 \times \1 \{X_{p+1} > 1\}$ and all the other conditional Kendall's tau are fixed as $\tau_{1,2|\X_J} = 0$.

\medskip

We randomly cut the sample into two parts of equal sizes ($500$ observations). The first part is used to construct the tree and therefore the partition of $\Rb^{d-p}$ using Algorithm~\ref{algo:tree_based_CKT}. The second part is used for the computation of the p-values of our tests. 
On average, between $2$ and $3$ boxes are selected for an average computation time between $0.2$ seconds (for $p = d-p = 1$) and $2.5$ seconds (for $p = d-p = 5$).
The estimated rejection probabilities are displayed on Table~\ref{tab:unknownBoxes_level} for the level analysis and on Table~\ref{tab:unknownBoxes_power} for the power analysis.

\medskip

We find that the powers and the levels of $\Tc_n^{(e)}$  (the test based on the covariance matrix) are rapidly degraded when the dimensions of the conditioning vector or of the conditioned vector increase. On the contrary, the bootstrap-based tests keep a good level and power in all cases. Note that, in general, this is not so much of a drawback since it is always possible to consider pairs of conditioned variables, so that a different statistic (and potentially a different partition tree) is computed for each couple of variables in the conditioned vector.

\medskip

\begin{table}[htb]
    \centering
    \begin{tabular}{|c|ccccc|ccccc|}
        \hline
        $d-p$ & 1 & 2 & 3 & 4 & 5
        & 1 & 2 & 3 & 4 & 5
        \\ \hline
        $p$ & \multicolumn{5}{c|}{Covariance matrix $\Tc_n^{(e)}$ }
        & \multicolumn{5}{c|}{NP Bootstrap $\Tc_{\infty,n}^{**}$}
        \\
        \hline

2 & 0.05 & 0.09 & 0.07 & 0.14 & 0.13
& 0.035 & 0.04 & 0.015 & 0.055 & 0.03 \\

3 & 0.24 & 0.275 & 0.36 & 0.255 & 0.3
& 0.05 & 0.025 & 0.04 & 0.065 & 0.04 \\

4 & 0.11 & 0.16 & 0.17 & 0.105 & 0.11
& 0.025 & 0.025 & 0.03 & 0.045 & 0.02 \\

5 & 0.13 & 0.16 & 0.18 & 0.135 & 0.185
& 0.025 & 0.015 & 0.035 & 0.02 & 0.045 \\

\hline




    \end{tabular}
    \caption{Empirical levels with unknown boxes for different dimensions of the conditioned and conditioning variables.}
    \label{tab:unknownBoxes_level}
\end{table}

\begin{table}[htb]
    \centering
    \begin{tabular}{|c|ccccc|ccccc|}
        \hline
        $d-p$ & 1 & 2 & 3 & 4 & 5 & 1 & 2 & 3 & 4 & 5
        \\ \hline
        $p$ & \multicolumn{5}{c|}{Covariance matrix $\Tc_n^{(e)}$ }
        & \multicolumn{5}{c|}{NP Bootstrap $\Tc_{\infty,n}^{**}$}
        \\

        \hline

2 & 1.00 & 0.99 & 0.995 & 1.00 & 0.995
& 1.00 & 1.00 & 1.00 & 1.00 & 1.00 \\

3 & 0.92 & 0.94 & 0.92 & 0.875 & 0.915
& 1.00 & 1.00 & 1.00 & 1.00 & 1.00 \\

4 & 0.835 & 0.835 & 0.805 & 0.785 & 0.84
& 0.995 & 0.995 & 0.995 & 0.995 & 1.00 \\

5 & 1.00 & 1.00 & 1.00 & 1.00 & 1.00
& 1.00 & 1.00 & 1.00 & 1.00 & 1.00 \\

        \hline

    \end{tabular}

    \caption{Empirical power with unknown boxes for different dimensions of the conditioned and the conditioning variables.
    }
    \label{tab:unknownBoxes_power}
\end{table}

\section{Two empirical applications}
\label{sect-applications}

\subsection{Financial dataset}

In this section, we consider time series of two stock indices, the Eurostoxx50 and the SP500. It is composed of $n=8265$ observations of daily returns from 5 January 1987 to 27 March 2020.
First, a preprocessing step is realized to get an ARMA-GARCH filtering of each of the marginal processes. 
The orders are selected by minimizing the BIC using the R package \texttt{fGarch}~\cite{fGarchpackage}. 
For the Eurostoxx (resp. SP500) returns, an ARMA(0,0)-GARCH(1,1) (resp. ARMA(1,1)-GARCH(1,1)) model is selected.
We denote by $X_{t,1}$ and $X_{2,t}$ the standardized residuals of the Eurostoxx and SP500 returns respectively, at time $t$.
The vector of interest will be therefore $\X_{t,I} := (X_{t,1}, X_{t,2})$.

\mds

First, we study the past residuals $\X_{t,J1} := L\X_{t,I} = (X_{t-1,1}, X_{t-1,2})$ denoting by $L$ the lag operator. 
Therefore, we can apply Algorithm~\ref{algo:tree_based_CKT}.
Using the classical nonparametric bootstrap test procedure, the differences between our estimated conditional Kendall's taus have been not significant (p-value $= 0.244$), and the assumption of a constant dependence structure cannot be rejected.


\mds

\begin{figure}[htb]
    \begin{center}
        \includegraphics[width=0.8\textwidth]{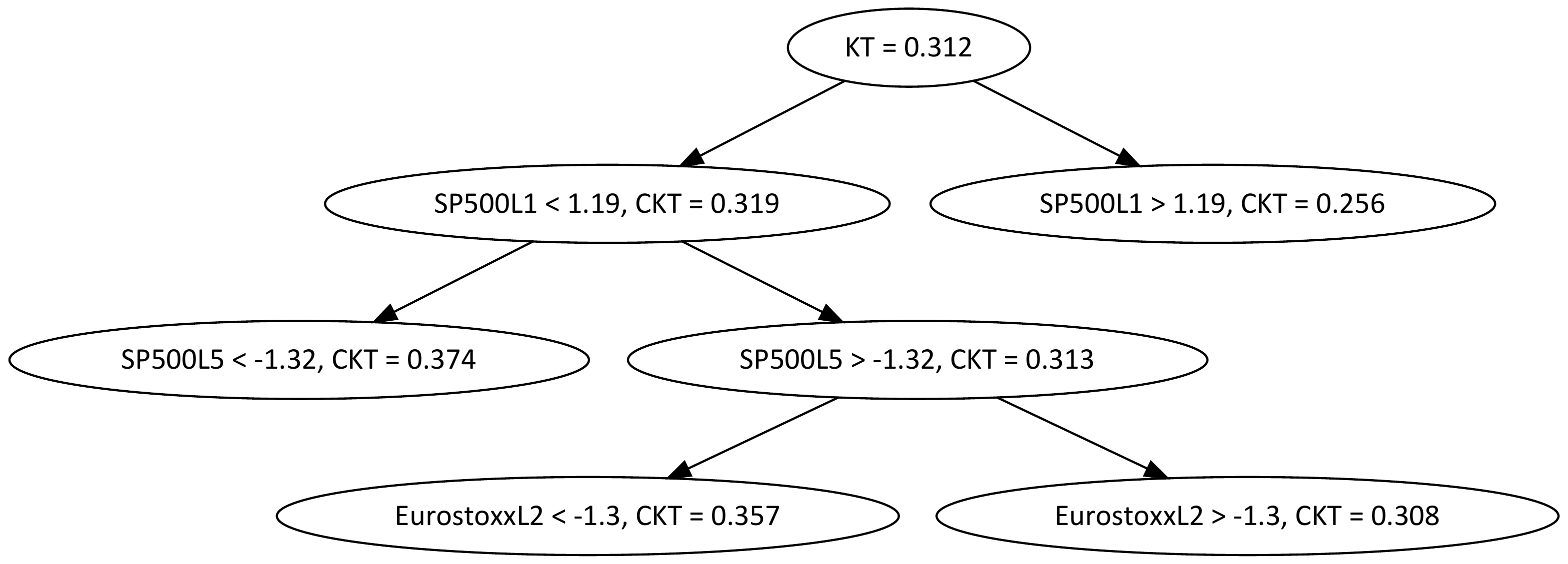}
    \end{center}
    \caption[Data-driven tree]{Data-driven tree for the conditional dependence between the Eurostoxx and SP500 innovations. Conditioning variables are their lagged innovations $\X_{t,J2}$ including up to $5$ lags.
    ``KT'' (resp. ``CKT'') denotes Kendall's tau (resp. conditional Kendall's tau at the event corresponding to the node).

    \smallskip

    Example: conditionally to $X_{t-1,2} < 1.19$ and $X_{t-5,2} > -1.32$, the conditional Kendall's tau between $X_{t,1}$ and $X_{t,2}$ is estimated as $\hat \tau_{X_{t,1}, \, X_{t,2} \,| \, X_{t-1,2} < 1.19, \, X_{t-5,2} > -1.32} = 0.313$.}
\label{fig:treeL5}
\end{figure}

We then decided to include more lags, using $\X_{t,J2} := (L\X_{t,I}, \dots, L^5\X_{t,I})$. Using Algorithm~\ref{algo:tree_based_CKT}, we obtain the tree displayed in Figure~\ref{fig:treeL5}. Contrary to the case of a single lag, this partition has been proved to be relevant: it induces significantly different conditional Kendall's taus, using the  classical nonparametric bootstrap test procedure (p-value $= 0$).

\mds

By a closer examination of the tree on Figure~\ref{fig:treeL5}, we can distinguish several regimes: in the normal regime, when the first lag of this innovation of the SP500 is greater than 1.19, the conditional Kendall's tau between the two innovations will be 0.256. Then we have an intermediate regime where SP500L1 $<$ 1.19, but EurostoxxL2 $>$ -1.3 and SP500L5 $>$ -1.32, where the conditional Kendall's tau is at 0.308. The conditional Kendall's tau for the left leaf is even higher at 0.357. The last case corresponds to the event in which the returns of SP500L1 and SP500L5 are both lower than usual. This surely represents a crisis-like situation and the conditional Kendall's tau reaches its maximum value 0.374.

\mds

It is interesting to note that, at each branch of the tree, the left-hand side node (corresponding to a lower lagged-innovation) has always a greater conditional Kendall's tau than the right-hand side node.
This illustrates the well-known contagion effect: when market conditions are bad, the dependencies between stock returns strengthen. Another way of seeing this contagion effect is by noticing that the tree displayed in Figure~\ref{fig:treeL5} is a \textbf{binary search tree}: every node has zero or two leaves and the value stored at every branch is smaller than the value on the left and bigger than the value on the right.
This means that, for every subset $A$ in our tree, every conditioning variable $X_k$ and every real $x$, we have
\begin{align}
    \tau_{1,2|\X_J \in A, \X_k \leq x}
    \geq \tau_{1,2|\X_J \in A}
    \geq \tau_{1,2|\X_J \in A, \X_k > x}.
    \label{eq:def_crisis_effect}
\end{align}
In other words, adding the information that $(X_k \leq x)$ leads to an increase in the dependence, and the opposite.
 
\bigskip

\begin{figure}[htb]
    \begin{center}
        \includegraphics[width=1.02\textwidth]{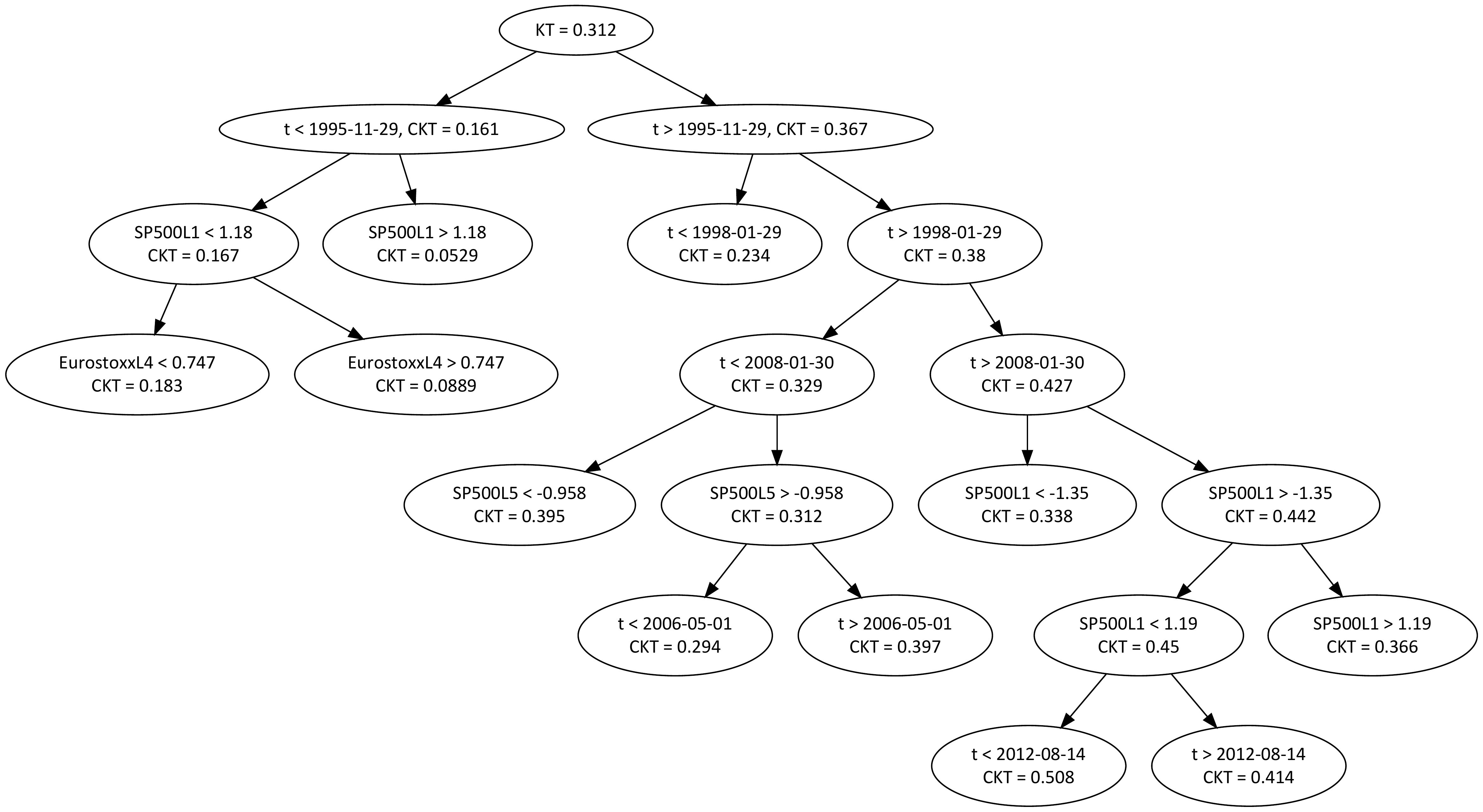}
    \end{center}
    \caption[Data-driven tree]{Data-driven tree for the conditional dependence between the Eurostoxx and SP500 innovations. Conditioning variables are their lagged innovations $\X_{t,J3}$ including up to $5$ lags and time.
    ``KT'' (resp. ``CKT'') denotes Kendall's tau (resp. conditional Kendall's tau at the event corresponding to the node).

    \smallskip

    Example: conditionally to $t <$ 1995-11-29 and $X_{t-1,2} < 1.18$, conditional Kendall's tau between $X_{t,1}$ and $X_{t,2}$ is estimated as
    $\hat \tau_{X_{t,1}, \, X_{t,2} \, | \,
    t < \text{1995-11-29}, \, X_{t-1,2} < 1.18} = 0.167$.}
\label{fig:treeL5t}
\end{figure}

In our last model, we decided to include also the time variable (even if it is not a random variable strictly speaking), so that $\X_{t,J
} := (L\X_{t,I}, \dots, L^5\X_{t,I}, t)$. This allows us to detect time-varying effects in the dependence between stock indices.
Using Algorithm~\ref{algo:tree_based_CKT}, we obtain the tree displayed in Figure~\ref{fig:treeL5t}. This partition induces strongly significant differences between Kendall's taus (p-value $= 0$, using the classical nonparametric bootstrap test procedures).
Three general effects can be noted:
\begin{itemize}
    \item The dependence between the SP500 and the Eurostoxx generally increases in time. This may be explained by the financial globalization in which major world indices become increasingly correlated with each other.
    \item The dependence between the SP500 and the Eurostoxx generally increases during periods with strongly negative innovations, compared to conditioning events where the stock indexes have positive innovations. This corresponds to the contagion effect (\ref{eq:def_crisis_effect}) that we have identified before.
    \item The latter contagion effect seems to be constant over time: Kendall's tau is around $0.10$ higher in the ``bad situations'' than it is in the ``good situations''.
    In other words, it means that $\tau_{1.2|\X_J \in bad, t \in T} - \tau_{1.2|\X_J \in good, t \in T}$ is nearly the same for each period of time.
\end{itemize}

\medskip

More precisely, we can see that the main split of the tree separates two periods of time. The second period is again split. In total, four main periods can be seen: the period 1987-1995 with a Kendall's tau of 0.161 between the SP500 and the Eurostoxx; then 1995-1998, 1998-2008 and 2008-2020, with the Kendall's tau of 0.234, 0.329 and 0.427 respectively. These division in the time variable illustrate the fact that the most important changes in the dependence between both financial indexes are linked to time trends and not to past returns. This is coherent with expected intuition: long-term phenomena such as globalization have more influence than short-term events such as stock return variations a few days before.
Note that the highest dependence levels are observed during the ``hot times'' 2008-2012, where financial markets suffered the financial crisis (Lehman's bankruptcy, in particular) followed by the European sovereign debt crisis 2010-2012.

\medskip

In the branch of the tree corresponding to the period 1987-1995, there are three leaves, corresponding to three conditioning subsets describing a bad situation, a good and an intermediate one, with increasing values of conditional Kendall's tau that satisfy~(\ref{eq:def_crisis_effect}).
The period 1995-1998 has not enough observations to be split further as it is already very short. It represents a transition (Kendall's tau = 0.234) between the previous period (Kendall's tau = 0.161) and the next one (Kendall's tau = 0.329).
In this period 1998-2008, Kendall's taus' are higher than before, but the branches still satisfy the general contamination principle~(\ref{eq:def_crisis_effect}).

\medskip

Interestingly, this principle is not satisfied in the most recent period (2008-2020), as the dependence during the ``stressed event'' $\{ X_{t-1,2} < -1.35 \}$ is in fact smaller (0.338) than during the complementary event (0.442).
It could be possible that such extreme events are linked to pure US news that do not affect so much the Eurostoxx. Nevertheless, in the ``normal'' branch $\{ X_{t-1,2} > -1.35 \}$, the classical behavior appears again, suggesting that the previous event $\{ X_{t-1,2} < -1.35 \}$ corresponds to a very special situation.

\FloatBarrier


\subsection{Insurance dataset}

In \cite{frees2016}, Frees, Lee and Yang have  presented an extensive analysis of an insurance dataset from  the Wisconsin Local Government Property Insurance Fund using  multivariate frequency-severity regression models. Their training sample covers the time period  from 2006 to 2010 and consists of  41 variables and 5677 observations. Each observation corresponds to a local government entity, which is a county, city, town, village, school and miscellaneous entity. Information on the type of a local entity, its number of claims and coverage sizes for a given year and insurance type, etc., are recorded by these 41 variables.

\medskip

\begin{figure}[p]
\begin{center}
\includegraphics[width=0.7\textwidth]{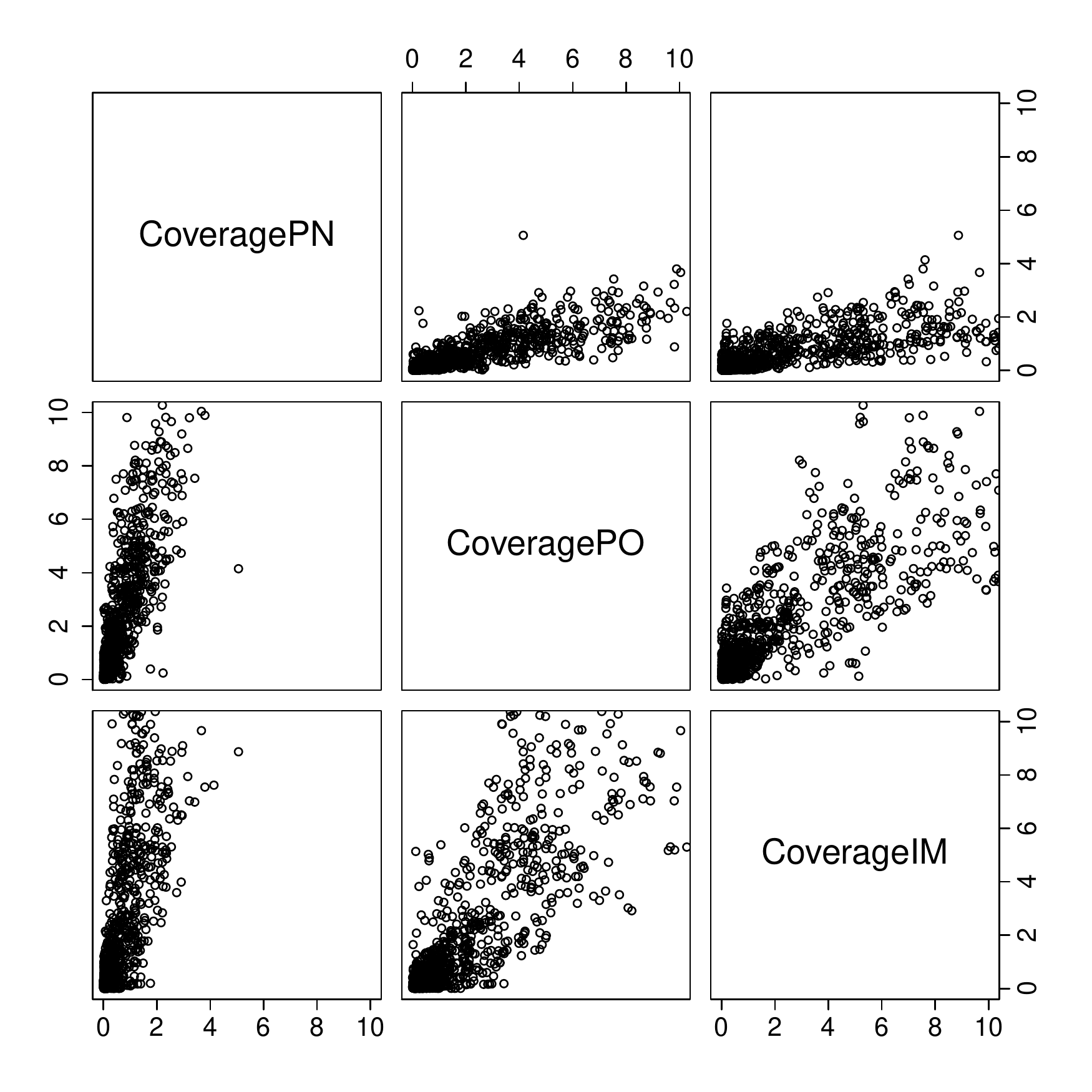}
\end{center}
\caption{Scatter plots of \texttt{CoveragePN}, \texttt{CoveragePO} and  \texttt{CoverageIM}}
 \label{fig-pairs-plot}
\end{figure}

The training data from \cite{frees2016} consists of nominal (\texttt{Type}), categorical (\texttt{Year}),  discrete and continuous variables. Continuous-type variables are non-negative and many of them  have an atom at 0.  We consider only the variables \texttt{Year} and  \texttt{Type}  as well as  the three continuous variables listed in Table \ref{tab-description-variables}. Further, we restrict ourselves to the observations for which the three claim coverages of interest have a positive logarithm. Further, the nominal variable \texttt{Type} classifies entities and  consists of six categories \texttt{City}, \texttt{County}, \texttt{School}, \texttt{Town}, \texttt{Village} and  \texttt{Miscellaneous}. We  exclude observations of miscellaneous entities to deal with entities of the same type within each of  five remaining  categories.
Thus, we finally obtain $1435$ observations. Figure \ref{fig-pairs-plot} displays the scatter plots of the three studied continuous variables, which are truncated for a better illustration. It should be noted that the conclusions of our statistical analysis hold only for  these three continuous variables knowing that the covered claims of a corresponding entity are larger than 1 million US dollars.

\medskip

\begin{table}[p]
\begin{center}
\begin{tabular}{|l|l|}\hline
\texttt{Variable} & Description \\ \hline \hline
\texttt{Year}    &    Claim year with values \texttt{2006}, \texttt{2007}, \texttt{2008}, \texttt{2009}, \texttt{2010}    \\ \hline
\texttt{Type}  &  Type of a local government entity with  nominal values  \\
   & \texttt{City}, \texttt{County}, \texttt{School}, \texttt{Town}, \texttt{Village},    \texttt{Miscellaneous}     \\ \hline
\texttt{CoveragePN}  &  Log coverage amount of comprehensive new vehicles (PN),  \\
             & where coverage is in millions of dollars (non-negative or null)  \\  \hline
\texttt{CoveragePO}   &   Log coverage amount of comprehensive old vehicles (PO), \\
             & where coverage is in millions of dollars (non-negative or null)    \\ \hline
\texttt{CoverageIM}   &   Log coverage amount of inland marine (IM), \\
             &  where coverage is in millions of dollars (non-negative or null)       \\ \hline
\end{tabular}
\caption{Description of variables}\label{tab-description-variables}
\end{center}
\end{table}

\begin{figure}[htb]
\begin{center}
\includegraphics[width=\textwidth]{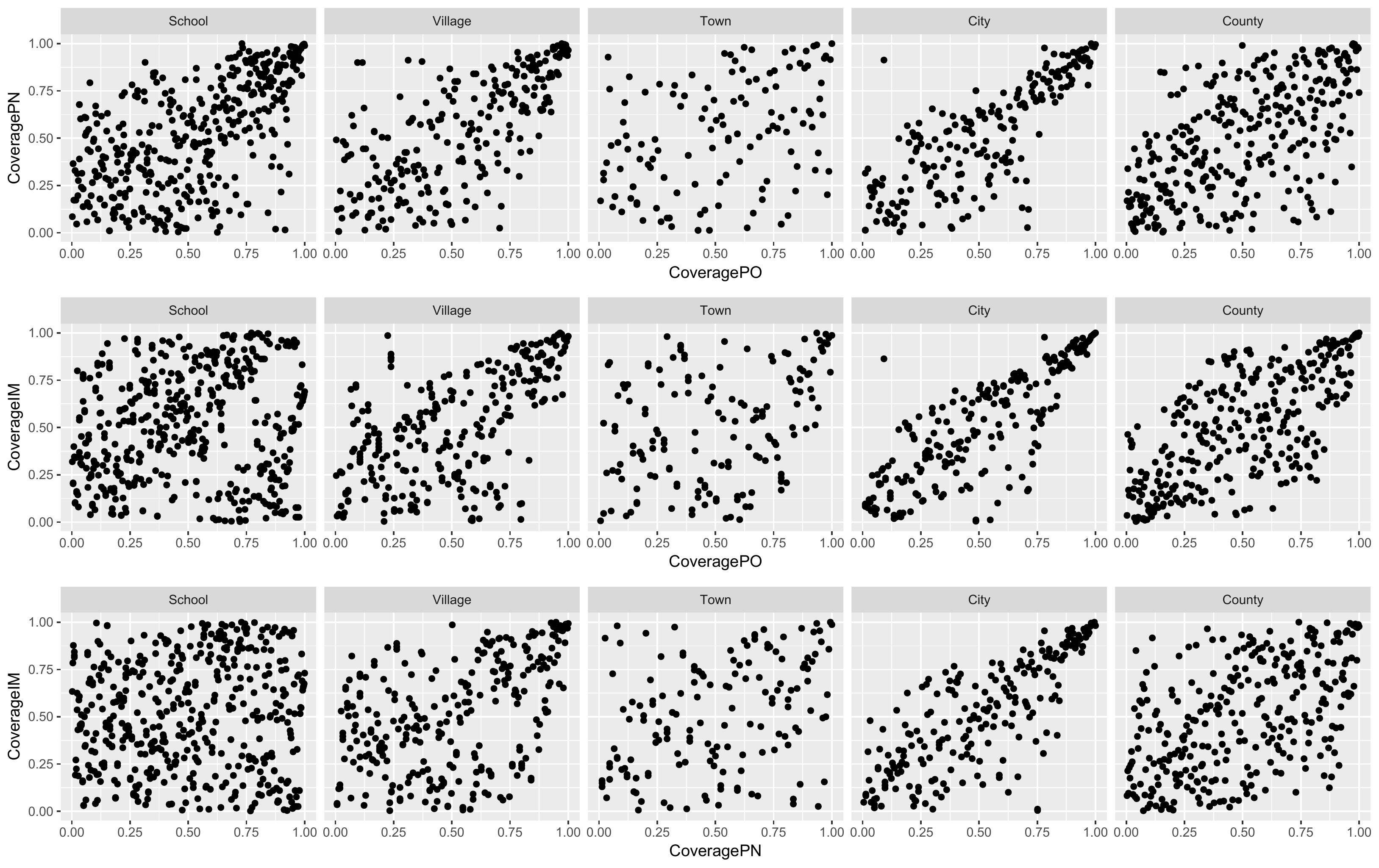}
\end{center}
\caption{Scatter plots on the unit square of \texttt{CoveragePN}, \texttt{CoveragePO} and  \texttt{CoverageIM} for each type of entity, when transformed to have (conditional) uniform margins.}
 \label{fig-kend-tau-different-entities}
\end{figure}

We apply the proposed  framework to the three latter continuous variables, to test whether their dependence structure varies across the considered five different years (2006, 2007, 2008, 2009, 2010).  Note that the conditioning variables specify boxes and they do not have to be continuous. The null hypothesis is that the three Kendall's taus between the variables \texttt{CoveragePN},  \texttt{CoveragePO} and   \texttt{CoverageIM} do not change over time. Under this null hypothesis, the test statistic $\Tc_n^{(e)}$ is $\chi^2-$distributed with $12$ degrees of freedom ($p=3$, five boxes), whose $95\%-$ quantile is  $21.026$. Taking the Kendall's tau for the year 2006 as the reference value and comparing it to the remaining Kendall's taus, i.e. we use the contrast matrix $T_e$ from \eqref{del_Te}, the  statistics $\Tc_n^{(e)}$ is equal to  $7.382$ (p-value=0.831) and   the null hypothesis cannot be rejected at $5\%$ significance level. 
Using 1000 bootstrap replicates of the data set, the four tests based on the bootstrapped test statistics cannot also reject the null hypothesis and their p-values range between 0.256 and 0.574.  

\mds

Now, we can pool data from different years together to test the null hypothesis $\bar\Hc_0^{\;\tau}$ of constant conditional dependence expressed by Kendall's taus  for the five different  entity types. Here, the null hypothesis can be rejected since the sample value of the test statistic  $\Tc_n^{(e)}$     is equal to $232.363$ (p-value=0). Using 1000 bootstrap replicates of the data set, the four tests based on the bootstrapped test statistics  also reject the null hypothesis and their p-values are  equal to 0.  Therefore, we can conclude that the  dependence structure of non-zero log coverage amounts  \texttt{CoveragePN}, \texttt{CoveragePO} and  \texttt{CoverageIM} should separately be modeled    for each entity type additionally to marginal modeling. Figure \ref{fig-kend-tau-different-entities} visualizes our conclusion. In order to exclude influence of conditional marginal distributions,  it  plots a copula data obtained from the original data using  (conditional) marginal empirical distribution functions for each box. Thus, one can  see in Figure \ref{fig-kend-tau-different-entities}  that  the dependence between   \texttt{CoverageIM} and \texttt{CoveragePO}   for the entity   \texttt{School} is  significantly lower than the same dependence for the entity  \texttt{City}. 


\section{Conclusion}

In this paper, we propose to test the assumption of constant conditional dependence for a set of several conditioning events using conditional Kendall's tau. Therefore, our testing approach is very simple, does not rely on the theory of empirical processes on the theoretical side, and has nice numerical performances. The asymptotic distribution of the considered Wald-type test statistics is a chi-square distribution, independently of any conditional marginal distribution. To avoid estimating a high-dimensional covariance matrix, we additionally consider two alternative test statistics, whose asymptotic distributions can be bootstrapped very fast. For this end, we promote the classical nonparametric bootstrap and the conditional bootstrap. In an extensive simulation study, we investigate empirical levels and power of the considered test statistics. An application to an insurance dataset illustrates the proposed test methods at work.

\medskip

In most applications, conditioning events are not known. We construct them ``blindly'' and recursively in such a way to maximize the difference between conditional Kendall's taus, adapting CART algorithm to the dependence framework. The output is a binary tree representing the decision paths to explain dependencies given some conditioning events. The leaves of the tree correspond to the final partition of conditioning events that are considered. Applying this to a dataset of financial returns, the estimated tree turns out to be a binary search tree, reflecting increasing dependencies during crises periods compared to non-crises ones.

\medskip

The proposed framework and ideas can be adapted to alternative dependence measures, as Spearman's rho. 
Moreover, several different multivariate dependence measures (\cite{SchmidMultivariateSpearman, SchmidConditionalSpearman, SchmmidEtAl2010, GNBG}, for instance) could be grouped to build richer and probably (even more) powerful test statistics.
This is a subject of an ongoing future research.

\section*{Acknowledgements}

Aleksey Min thanks Jae Youn Ahn for pointing out the insurance data from Wisconsin Local Goverment Property Insurance Fund.
Jean-David Fermanian's work has been supported by the labex Ecodec (reference project ANR-11-LABEX-0047).

\bigskip

\newpage

\appendix

\section{Proofs}
\label{sect-proofs}

\subsection{Proof of Theorem~\ref{AN_hatV}} \label{Proof-Theorem-AN_hatV}
Consider a deterministic vector $\a \in \Rb^{2m}$. Once proved the asymptotic normality of the random variable $\sqrt{n}\,\a' \hat V$,
the weak convergence of $\sqrt{n}\,\hat V$ will be obtained by the usual Cramer-Wold device.

\mds
Before, we approximate the U-statistics $\hat D_k$ by its Haj\`ek projection. To this goal, let us symmetrize the latter quantities:
$$ \hat D_k := \frac{1}{2n(n-1)} \sum_{i=1}^n \sum_{j=1, j\neq i}^n (g_{ij,k} + g_{ji,k}),$$
$$ g_{ij,k}:=\1\{ X_{i,1}< X_{j,1}, X_{i,2}< X_{j,2},X_{i,J}\in A_{k,J}, X_{j,J}\in A_{k,J} \}.$$
Set $g^*_{ij,k}:=(g_{ij,k} + g_{ji,k})/2$ and
$ \tilde D_k :=2n^{-1} \sum_{i=1}^n \Eb\big[ g^*_{i0,k}  | \X_i \big]-D_k ,$
introducing another independent realization $\X_0$.
Note that
$$\Eb[\tilde D_k]= \Eb\big[ 2g^*_{i0,k}] - D_k=2\Eb[ g_{1,2,k}]- D_k =D_k.$$
Moreover, simple calculations yield
\begin{eqnarray*}
\lefteqn{ \hat D_k - \tilde D_k = \frac{1}{n(n-1)}\sum_{i\neq j} g^*_{ij,k} - \frac{1}{n}\sum_{i=1}^n \Eb[g^*_{i0,k} | \X_i]  - \frac{1}{n}\sum_{i=1}^n  \Eb[g^*_{0i,k} | \X_i]+D_k    }\\
&=&\frac{1}{n(n-1)}\sum_{i\neq j} \big(g^*_{ij,k} - \Eb[g^*_{ij,k} | \X_i] - \Eb[g^*_{ji,k} | \X_i]+D_k   \big) \\
&=:& \frac{1}{n(n-1)}\sum_{i\neq j} \bar g_{ij,k}.\hspace{6cm}
\end{eqnarray*}
Note that $\Eb[\bar g_{ij,k}| \X_i]=\Eb[\bar g_{ij,k}| \X_j]=0$.
By a usual reasoning for U-statistics, it may be easily checked that
$$ \var\big( \hat D_k - \tilde D_k \big) =
\frac{1}{n^2(n-1)^2}\sum_{i_1\neq j_1} \sum_{i_2\neq j_2} \Eb\big[ \bar g_{i_1 j_1,k} \bar g_{i_2 j_2,k} \big] = O\big(n^{-2}\big).$$
Indeed, the previous cross-products are zeros when some of the four indices $(i_1,j_1,i_2,j_2)$ is different from the others.
This yields $\hat D_k=\tilde D_k + O_P(n^{-1})$ and we deduce
\begin{eqnarray*}
\lefteqn{ \sqrt{n}\,\a' \hat V=
\sqrt{n}\sum_{k=1}^m a_k (\tilde D_k- D_k)+\sqrt{n} \sum_{k=m+1}^{2m} a_k (\hat p_k - p_k)+O_P(n^{-1/2}) }\\
&=&  n^{-1/2} \sum_{i=1}^n \Big\{ \sum_{k=1}^m 2 a_k \big( \Eb\big[ g^*_{i0,k}  | \X_i \big]-D_k \big) +
\sum_{k=m+1}^{2m} a_k \big( \1\{ X_{i,J}\in A_k\} - p_k  \big) \Big\} +O_P(n^{-1/2}) \\
&=& n^{-1/2} \sum_{i=1}^n \a' \tilde \vv_i + O_P(n^{-1/2}),
\end{eqnarray*}
denoting by $\tilde \vv_i$ the column random vector
$$ \tilde \vv_i=\Big[ 2 \big( \Eb[ g^*_{i0,1}  | \X_i ]-D_1 \big),\ldots, 2 \big( \Eb[ g^*_{i0,m}  | \X_i ]-D_m \big),\1\{ X_{i,J}\in A_1\} - p_1,\ldots,\1\{ X_{i,J}\in A_m\} - p_m  \Big]'.$$
By the usual CLT, we deduce that $\sqrt{n}\,\a' \hat V$ tends in law towards a $\Nc(0,\a' \Sigma \a)$, $\Sigma=\Eb[ \tilde \vv_i' \tilde \vv_i]$.
Since this is true for every vector $\a$, this means
$\sqrt{n}\,\hat V \leadsto \Nc(0,\Sigma)$.
Note that
\begin{eqnarray*}
\lefteqn{ \Eb\big[ g^*_{i0,k}  | \X_i \big]= \1\{\X_{i,J}\in A_{k,J}\}\int \pi(\x,\X_i) \1\{\x_J \in A_{k,J}\}\, \Pb(d\x) }\\
&=& p_k \1\{\X_{i,J}\in A_{k,J}\}\int \pi(\x,\X_i) \, \Pb_k(d\x).
\end{eqnarray*}
Therefore, simple calculations yield the components of $\Sigma$.
For example,
\begin{eqnarray*}
\sigma_{k,l}&:=& 4\Eb\Big[\Eb\big[ g^*_{i0,k}  | \X_i \big]\Eb\big[ g^*_{i0,l}  | \X_i \big] \Big]-  4 D_k D_l\\
&=&4 p_k p_l \Eb\Big[ \1\{\X_{i,J}\in A_{k,J},\X_{i,J}\in A_{l,J}\} \int \pi(\x_1,\X_i) \, \Pb_k(d\x_1)\int \pi_l(\x_2,\X_i) \, \Pb_l(d\x_2) \Big]-  4 D_k D_l \, .
\end{eqnarray*}

If $k = l$, note that
\begin{eqnarray*}
\sigma_{k,k}&:=& 4p_k^2\Eb\Big[ \1 \{\X_{i,J}\in A_{k,J}\}\int \pi(\x_1,\X_i) \, \Pb(d\x_1)\int \pi(\x_2,\X_i) \, \Pb(d\x_2)\Big] -  4 D_k^2\\
&=& 4 p_k^3\int_{\x_3 \in A_{k,J}}\Big[\int \pi(\x_1,\x_3) \, \Pb_k(d\x_1)\int \pi(\x_2,\x_3) \, \Pb_k(d\x_2) \Big] \Pb_k(d\x_3) -  4 D_k^2.
\end{eqnarray*}

Concerning $\Sigma_{1,2}:=[\rho_{k,l}]_{1\leq k,l \leq m}$, we get
\begin{eqnarray*}
\rho_{k,l}&:=& 2\Eb\Big[\Eb\big[ g^*_{i0,k}  | \X_i \big] \1\{X_{i,J}\in A_{l,J} \} \Big]-  2 D_k p_l\\
&=& 2p_k \Eb\Big[ \1\{\X_{i,J}\in A_{k,J} \cap A_{l,J}\} \int \pi(\x,\X_i) \, \Pb_k(d\x) \Big]-  2 D_k p_l \, .
\end{eqnarray*}
When $k = l$, we have
\begin{eqnarray*}
\rho_{k,k}&:=& 2p_k\Eb\Big[ \1\{\X_{i,J}\in A_{k,J}\}\int \pi(\x,\X_i) \, \Pb_k(d\x)\Big] -  2 D_k p_k\\
&=& 2p^2_k\int\Big[\int \pi(\x_1,\x_2) \, \Pb_k(d\x_1) \Big] \Pb_k(d\x_2) -  2 D_k p_k \\
&=& 2D_k -  2 D_k p_k.
\end{eqnarray*}

\subsection{Proof of Proposition~\ref{AN_hatW}} \label{Proof-Propo-2}

Let us prove the asymptotic normality of $n^{1/2}\hat W^{(1)}$ and the result will follow.
Note that, for every $k\in \{1,\ldots,m\}$, the $k$-th component of $\hat W^{(1)}$ is
\begin{eqnarray*}
\lefteqn{ \sqrt{n}( \hat \tau^{(1)}_{1,2|\X_J \in A_{k,J} } - \tau_{1,2|\X_J \in A_{k,J} })
= 4\sqrt{n}\Big(  \frac{\hat D_k}{\hat p_k^2} - \frac{ D_k}{p_k^2} \Big)     }\\
&=&
4\sqrt{n}\Big(  \frac{\hat D_k- D_k}{p_k^2}\big( 1 + \frac{p_k^2 - \hat p_k^2}{\hat p_k^2}\big) + \frac{ D_k \big( p_k^2 - \hat p_k^2 \big)}{\hat p_k^2 p_k^2} \Big)   \\
&=&
4\sqrt{n}\Big(  \frac{\hat D_k- D_k}{p_k^2} + O_P\big( (\hat D_k - D_k) (\hat p_k - p_k)\big)  - \frac{ 2 D_k  (\hat p_k - p_k)}{\hat p_k^2 p_k}  - \frac{ D_k  (\hat p_k - p_k)^2}{\hat p_k^2 p_k^2} \Big)   \\
&=&
4\sqrt{n}\Big(  \frac{\hat D_k- D_k}{p_k^2} - \frac{2 D_k \big( \hat p_k - p_k \big)}{ p_k^3} \Big) +O_P(n^{-1/2})=: 4 \zeta_k + O_P(n^{-1/2}),
\end{eqnarray*}
due to Theorem~\ref{AN_hatV}.
With the notations of $\Sigma$'s components, direct calculations provide, for every $k,l$ in $\{1,\ldots,m\}$,
\begin{eqnarray*}
\lefteqn{ \Eb[\zeta_k \zeta_l]=
\frac{\sigma_{kl}}{p_k^2p_l^2} + \frac{4 D_k D_l \big( p_{k,l}- p_k p_l \big)}{ p_k^3 p_l^3} }\\
&-& \frac{2 D_l \big( 2p_k J_{k,l}- 2D_k p_l \big)}{ p_k^2 p_l^3}
- \frac{2 D_k \big( 2p_l J_{l,k}- 2D_l p_k \big)}{ p_l^2 p_k^3}    \\
&=&
\frac{4 I_{k,l}}{p_k^2p_l^2} + \frac{4 D_k D_l p_{k,l}}{ p_k^3 p_l^3}
- \frac{4 D_l J_{k,l}}{ p_k p_l^3} - \frac{4 D_k J_{l,k}}{ p_l p_k^3}\cdot
\end{eqnarray*}
Since $4D_k/p_k^2 = 1+ \tau_{1,2|\X_J \in A_{k,J}}$, we get the result when $k=l$.

\subsection{Proof of Proposition~\ref{AN_hatV_extended}}

Consider a deterministic vector $\a \in \Rb^{m+p(p-1)m/2}$. It will be decomposed as a block column vector
$$ \a := [\a'_0,\a'_{1,2},\a'_{1,3},\ldots,\a'_{p-1,p}]',$$
$$ \a_{0}:= [a_{0,1},\ldots,a_{0,m}],\;\; \a_{a,b}:=[a_{a,b,1},\ldots,a_{a,b,m}].$$

By proving the asymptotic normality of the random variable $\sqrt{n}\,\a' \hat V$,
the weak convergence of $\sqrt{n}\,\hat V$ will be obtained by invoking the usual Cramer-Wold device.
As in the proof of Theorem~\ref{AN_hatV}, for any couple $(a,b)$, set
$$ \hat D_{a,b,k} := \frac{1}{2n(n-1)} \sum_{i=1}^n \sum_{j=1, j\neq i}^n \big(g_{ij,abk} + g_{ji,abk}\big),$$
$$ g_{ij,abk}:=\1\{ X_{i,a}< X_{j,a}, X_{i,b}< X_{j,b},X_{i,J}\in A_{k,J}, X_{j,J}\in A_{k,J} \}.$$
Set $g^*_{ij,abk}:=(g_{ij,abk} + g_{ji,abk})/2$ and
$ \tilde D_{a,b,k} :=2n^{-1} \sum_{i=1}^n \Eb\big[ g^*_{i0,abk}  | \X_i \big]-D_{a,b,k} .$
Obviously, $\Eb[\tilde D_{a,b,k}]=D_{a,b,k}.$
Note that $$ \Eb\big[ g^*_{i0,abk}  | \X_i \big]= p_k \1\{\X_{i,J}\in A_{k,J}\}\int \pi_{abk}(\x,\X_i) \, \Pb_k(d\x).$$
We get $\hat D_{a,b,k}=\tilde D_{a,b,k} + O_P(n^{-1})$ by a usual U-statistics reasoning, and we deduce
\begin{eqnarray*}
\lefteqn{ \sqrt{n}\,\a' \hat \V=
\sqrt{n}\sum_{(a,b)}\sum_{k=1}^m a_{a,b,k} (\tilde D_{a,b,k}- D_{a,b,k})+\sqrt{n} \sum_{k=1}^{m} a_{0,k} (\hat p_k - p_k)+O_P(n^{-1/2}) }\\
&=&  n^{-1/2} \sum_{(a,b)} \sum_{i=1}^n \Big\{ \sum_{k=1}^m 2 a_{a,b,k} \big( \Eb\big[ g^*_{i0,abk}  | \X_i \big]-D_{a,b,k} \big) +
\sum_{k=1}^{m} a_{0,k} \big( \1\{ X_{i,J}\in A_k\} - p_k  \big) \Big\} +O_P(n^{-1/2}) \\
&=& n^{-1/2} \sum_{i=1}^n \a' \tilde \V_i + O_P(n^{-1/2}),
\end{eqnarray*}
denoting by $\tilde \V_i$ the column random vector
$$ \tilde \V_i=\big[ \tilde \V'_{i,0},\tilde \V'_{i,1,2},\tilde \V'_{i,1,3},\ldots,\tilde \V'_{i,p-1,p} \big]',$$
$$ \tilde \V_{i,a,b}=\big[ 2 \big( \Eb[ g^*_{i0,ab1}  | \X_i ]-D_{a,b,k} \big),\ldots, 2 \big( \Eb[ g^*_{i0,abm}  | \X_i ]-D_{a,b,m}) \big]',$$
$$ \tilde \V_{i,0} :=\big[ \1\{ X_{i,J}\in A_1\} - p_1,\ldots,\1\{ X_{i,J}\in A_m\} - p_m  \big]'.$$
By the usual CLT, we deduce that $\sqrt{n}\,\a' \hat \V$ tends in law towards a $\Nc(0,\a' \Sigma \a)$, $\Sigma=\Eb[ \tilde \V_i' \tilde \V_i]$.
Since this is true for every vector $\a$, this means
$\sqrt{n}\,\hat \V \leadsto \Nc(0,\Sigma_e)$.
The calculation of $\Sigma_e$ is similar to the calculations in the proof of Theorem~\ref{AN_hatV}.

\subsection{Proof of Proposition~\ref{AN_hatW_extended}}

Let us prove the asymptotic normality of $n^{1/2}\hat \W^{(1)}$ and the result will follow. As in the proof of Proposition~\ref{AN_hatW}, for every couple $(a,b)\in \{1,\ldots,p\}^2$, $a\neq b$, and every $k\in \{1,\ldots,m\}$, we have
\begin{eqnarray*}
\lefteqn{ \sqrt{n}( \hat \tau^{(1)}_{a,b|\X_J \in A_{k,J} } - \tau_{a,b|\X_J \in A_{k,J} })
= 4\sqrt{n}\Big(  \frac{\hat D_{a,b,k}}{\hat p_k^2} - \frac{ D_{a,b,k}}{p_k^2} \Big)     }\\
&=&
4\sqrt{n}\Big(  \frac{\hat D_{a,b,k}- D_{a,b,k}}{p_k^2} - \frac{2 D_{a,b,k} \big( \hat p_k - p_k \big)}{ p_k^3} \Big) +O_P(n^{-1/2})=: 4 \zeta_{a,b,k} + O_P(n^{-1/2}),
\end{eqnarray*}
due to Theorem~\ref{AN_hatV_extended}.
Direct calculations provide, for every couples $(a,b)$, $(a',b')$ and every $k,l$ in $\{1,\ldots,m\}$,
\begin{eqnarray*}
\lefteqn{ \Eb[\zeta_{a,b,k} \zeta_{a',b',l}]=
\frac{4 p_k p_l I_{a,b,a',b',k}  - 4 D_{a,b,k} D_{a',b',l}}{p_k^2p_l^2} + \frac{4 D_{a,b,k} D_{a',b',l} \big( p_{k,l}- p_k p_l \big)}{ p_k^3 p_l^3} }\\
&-& \frac{2 D_{a',b',l} \big( 2p_k J_{a,b,k,l}- 2D_{a,b,k} p_l \big)}{ p_k^2 p_l^3}
- \frac{2 D_{a,b,k} \big( 2p_l J_{a',b',l,k} - 2D_{a',b',l}  p_k \big)}{ p_l^2 p_k^3}    \\
&=& 
4\Big( \frac{ I_{a,b,a',b',k,l}}{p_k p_l} + \frac{ D_{a,b,k} D_{a',b',l} p_{k,l}}{ p_k^3 p_l^3}
- \frac{ D_{a',b',l} J_{a,b,k,l}}{ p_k p_l^3} -  \frac{ D_{a,b,k} J_{a',b',l,k}}{ p_l p_k^3} \Big),
\end{eqnarray*}
yielding the announced result.

\section{Two counter-examples}
\label{sec:no_link_CKT}
In this section, we inspect the (lack of) relationships between pointwise conditioning and subset conditioning events for conditional Kendall's tau.
We introduce the two trivariate following models. 

\mds

1. Set $I=\{1,2\}$, $J=\{3\}$. Assume $X_3$ follows a uniform distribution on $[0,4]$.
For every $x_3 \in [0,4]$:
\begin{itemize}
    \item given $X_3=x_3 \in [0,1]$, $X_1$ and $X_2$ are two independent uniform distributions on $[0,1]$;
    \item given $X_3=x_3 \in [1,2]$, $X_1$ and $X_2$ are two independent uniform distributions on $[2,3]$;
    \item given $X_3=x_3 \in [2,3]$, $X_1$ and $X_2$ are two independent uniform distributions respectively on $[0,1]$ and $[2,3]$;
    \item given $X_3=x_3 \in [3,4]$, $X_1$ and $X_2$ are two independent uniform distributions respectively on $[2,3]$ and $[0,1]$.
\end{itemize}
For every $x_3 \in [0,4]$, $\tau_{1,2|\X_J=\x_J} = 0$ by independence. We choose $A_1 = [0,2]$ and $A_2 = [2,4]$.
Then, simple calculations shows that $\tau_{1,2|\X_J \in A_1} = 1/2$ and $\tau_{1,2|\X_J \in A_2} = -1/2$ (calculate $p_{1,2|A_k}$ and consider the locations of $X_1$ and $X_2$ in different intervals). This provides a interesting ``counter-example'': in this model,
$\x_J \mapsto \tau_{1,2|\X_J=\x_J}$ is constant but $A \mapsto \tau_{1,2|\X_J \in A}$ is not constant.

\bigskip

2. Set $I=\{1,2\}$, $J=\{3\}$. Assume $X_3$ follows a uniform distribution on $[0,4]$.
For every $x_3 \in [0,4]$:
\begin{itemize}
    \item given $x_3 \in [0,1]$, $(X_1,X_2)$ follows a Gaussian copula with correlation $1/2$;
    \item given $x_3 \in [1,2]$, $(X_1,X_2)$ follows a Gaussian copula with correlation $-1/2$;
    \item given $x_3 \in [2,3]$, $(X_1,X_2)$ follows a Gaussian copula with correlation $1/2$;
    \item given $x_3 \in [3,4]$, $(X_1,X_2)$ follows a Gaussian copula with correlation $-1/2$.
\end{itemize}
By construction, the function $\x_J \mapsto \tau_{1,2|\X_J=\x_J}$ is not constant, because the correlation of $(X_1,X_2)$ is a non constant function of $x_3$.
Set $A_1 = [0,2]$ and $A_2 = [2,4]$.
The function $A \mapsto \tau_{1,2|\X_J \in A}$ is constant because
the laws of $(X_1, X_2)$ knowing $X_3 \in A_1$ and of $(X_1, X_2)$ knowing $X_3 \in A_2$ are equal. Indeed, they are both mixtures of two Gaussian copulas, with weights $(1/2,1/2)$ and correlations $(1/2, -1/2)$.
This provides a counter-example where
$\x_J \mapsto \tau_{1,2|\X_J=\x_J}$ is not constant
but $A \mapsto \tau_{1,2|\X_J \in A}$ is constant.

\end{document}